\documentclass[onecolumn]{mn2e}
\usepackage{psfig}
\newcommand{\beq}{\begin{equation}}
\newcommand{\eeq}{\end{equation}}
\newcommand{\bdm}{\begin{displaymath}}
\newcommand{\edm}{\end{displaymath}}
\newcommand{\bfig}{\begin{figure}}
\newcommand{\efig}{\end{figure}}
\newcommand{\msun}{M_{\odot}}
\def\etal{{\it et al.~}}
\def\ie{{\frenchspacing\it i.e. }}
\def\eg{{\frenchspacing\it e.g. }}

\def\gtsima{$\; \buildrel > \over \sim \;$}
\def\ltsima{$\; \buildrel < \over \sim \;$}
\def\prosima{$\; \buildrel \propto \over \sim \;$}
\def\gsim{\lower.5ex\hbox{\gtsima}}
\def\lsim{\lower.5ex\hbox{\ltsima}}
\def\simgt{\lower.5ex\hbox{\gtsima}}
\def\simlt{\lower.5ex\hbox{\ltsima}}
\def\simpr{\lower.5ex\hbox{\prosima}}

\def\HH{H$_2$~}

\title[Primordial Fragmentation]{Fragmentation and the formation of
primordial protostars:
the possible role of Collision Induced Emission}

\author[E. Ripamonti and T. Abel]{E. Ripamonti$^{1}$\thanks{E-mail:
ripa@astro.psu.edu (ER); hi@tomabel.com (TA)} and
T. Abel$^{1}$\footnotemark[1]\\
$^{1}$ Pennsylvania State University, Dept. of Astronomy and 
Astrophysics, 525 Davey Lab, University Park, PA 16802, U.S.A.}
\begin{document}

\date{Submitted: 2003 September 4}

\pagerange{\pageref{firstpage}--\pageref{lastpage}} \pubyear{2003}

\maketitle

\label{firstpage}

\begin{abstract}
The mechanisms which could lead to chemo-thermal instabilities and
fragmentation during the formation of primordial protostars are
investigated analytically. We introduce new analytic approximations for
\HH cooling rates bridging the optically thin and thick regimes. These
allow us to discuss chemo--thermal instabilities up to densities when
protostars become optically thick to continuum radiation
($n\equiv\rho/m_H\lsim10^{16-17}\;{\rm cm^{-3}}$).  During the
proto-stellar collapse instabilities are active in two different density
regimes. In the well known ``low density'' regime
($n\sim10^8-10^{10}\;{\rm cm^{-3}}$), instability is due to 3-body
reactions quickly converting atomic hydrogen into H$_2$. In the ``high
density'' regime ($n\gsim 10^{14}\;{\rm cm^{-3}}$), another instability
is triggered by the strong increase in the cooling rate due to \HH
Collisional Induced Emission (CIE). In agreement with the three
dimensional simulations, we find that the ``low density'' instabilities
cannot lead to fragmentation, both because fluctuations are too small to
survive turbulent mixing, and because their growth times are too
slow. The situation for the newly found ``high density'' instability is
analytically similar. This continuum cooling instability is as weak as
``low density'' instability, with similar ratios of growth and dynamical
time scales, as well as allowing for the necessary fragmentation
condition $t_{cool}\lsim t_{dyn}$. Because the instability growth
timescale is always longer than the free fall timescale, it seems
unlikely that fragmentation could occur in this high density regime.
Consequently, one expects the first stars to be very massive, not to
form binaries nor harbour planets.  Nevertheless, full three dimensional
simulations are required to be certain.  Such 3D calculations could
become possible using simplified approaches to approximate the effects of
radiative transfer, which we show to work very well in 1D calculations,
giving virtually indistinguishable results from calculations employing
full line transfer. This indicates that the effects of radiative
transfer during the initial stages of formation of primordial
proto--stars are local corrections to cooling rather than influencing
the energetics of distant regions of the flow.
\end{abstract}


\section{Introduction}

In current scenarios for the formation of the first stars, molecular
hydrogen plays a prominent role, providing the only cooling mechanism
for metal-free gas at $T\lsim 10^4\;{\rm K}$. \HH cooling is
responsible for both the collapse of the gas inside the small ($M_{\rm
halo} \sim 10^6\;\msun$) cosmological halos where primordial star
formation is believed to occur (\eg Couchman \& Rees 1986; Tegmark
\etal 1997, Abel \etal 1998), and for the fragmentation of the gas
itself into ``clumps'' with masses of $\sim 100-1000\;\msun$, as is
seen in the simulations of Abel, Bryan \& Norman (1998, 2000) (see
also Bromm, Coppi \& Larson 2002). The fate of these objects is
unclear. Several previous studies, based on analytical arguments (\ie
stability analysis) or single-zone models, led to different
conclusions about the fragmentation properties of primordial clouds.
One important instability-triggering mechanism was first suggested by
Palla, Salpeter \& Stahler (1983; hereafter PSS83): the onset of
3-body \HH formation at number densities $n\gsim 10^8\;{\rm cm^{-3}}$
(where $n\equiv\rho/m_{\rm H}$) causes a fast increase in the cooling
rate, possibly leading to fragmentation on a mass scale of $\sim
0.1\msun$. A similar result was obtained by Silk (1983; hereafter S83)
by applying a criterion for ``chemo-thermal instability'' (first
discussed by Sabano \& Yoshii, 1977) to an object where 3-body \HH
formation is important.

Numerical simulations have recently shed new light upon the issue, and
in particular Abel, Bryan \& Norman (2002; hereafter ABN02) were able to
reach the 3-body reactions density regime finding that these reactions
actually lead to the formation of a single fully molecular core (with a
mass of $\sim 1\;\msun$) at the centre of each clump. They point out
that in their simulations they see several chemo-thermally unstable
regions, but that the instabilities do not lead to fragmentation because
turbulence efficiently mixes the gas, erasing each fluctuation before it
can grow significantly. Recently, Omukai \& Yoshii (2003; OY03) found a
somewhat different explanation by improving the original S83
investigation and pointing out that instabilities growth was too slow.

The further evolution of such an object is still uncertain:
due to the high molecular fraction, many of the numerous \HH
roto-vibrational lines (accounting for most of the cooling) become
optically thick just after the formation of the molecular core. For this
reason, it becomes impossible to estimate the \HH cooling rate without a
complete treatment of radiative transfer, which currently can not be
included into full 3-D hydrodynamical simulations. Simpler 1-D studies,
such as those in Omukai \& Nishi (1998; hereafter ON98) and Ripamonti
\etal (2002; hereafter R02), can follow the further evolution, giving useful
predictions about the physical processes driving the collapse, or about
the properties of the small hydrostatic protostellar core which finally
forms in the centre; on the other hand, their intrinsic spherical
symmetry prevents them from giving direct predictions about
fragmentation.

An interesting result of ON98 and R02 is that during the collapse, the
protostellar object experiences a phase when cooling is dominated by
the effects of \HH Collision-Induced Emission (CIE), the opposite
process of the more commonly known Collision-Induced Absorption, or
CIA. Both of them find that, at the beginning of the
CIE-cooling phase, the molecular core is optically thin to
CIE-produced continuum photons, and there is a very fast increase in
the cooling rate until the continuum optical depth exceeds
unity. However, neither of these previous studies pointed out that
during this phase the core conditions resemble the ones that, at the
onset of 3-body \HH formation, led to the chemo-thermal instability:
the core is optically thin (which is believed to be a necessary
condition for fragmentation; see Rees 1976) and the cooling rate
is undergoing a dramatic increase.

In the next sections, we will investigate more thoroughly the
issue of chemo-thermal instability, with special attention to the
effects of CIE cooling. In section 2 we will describe the main cooling
processes, giving a brief account of CIE emission, and some useful
approximations for \HH line cooling. In section 3 we examine the
conditions for the insurgence of chemo-thermal instabilities, and
argue whether they can cause fragmentation. Finally, we discuss the
results in section 4.

\section{Cooling processes}

\subsection{Collision-Induced Emission}

\HH molecules have no electric dipole, and emission or absorption of
radiation can take place only through quadrupole transitions. But when
a collision takes place, the interacting pair (H$_2$-H$_2$, H$_2$-He,
H$_2$-H) briefly acts as a ``supermolecule'' with a nonzero electric
dipole, and an high probability of emitting (CIE) or absorbing (CIA) a
photon (see the brief discussion in Lenzuni, Chernoff \& Salpeter
1991, or Frommhold 1993 for an extensive account). In the same way,
during an H-He collision, the two atoms perturb each other and there
is a relevant probability of emission (or absorption) of a photon through
a dipole transition. Because of the very short collision times
($\Delta t\lsim10^{-12}\;{\rm s}$ for $T\gsim 300\;{\rm K}$),
collision-induced lines become very broad, actually merging into a
continuum; for example, in the \HH CIE spectrum only the vibrational
bands can be discerned as smooth peaks deriving from the merging of
the roto-vibrational lines.

\subsubsection{CIE cooling rate}

\begin{table}
\caption{List of references for the various kind of collisions leading
to CIE emission. We also specify the temperature range and the maximum
frequency considered.}
\label{cie_references}
\begin{tabular}{ccccl}
\hline\hline
Pair & T[K] & $\nu_{\rm max}$[cm$^{-1}$] & Reference\\
\hline
H$_2$-\HH & 400-1000   & 17000 & Borysow 2002 \\
H$_2$-\HH & 1000-7000  & 20000 & Borysow \etal 2001 \\
H$_2$-He  & 1000-7000  & 20088 & Jorgensen \etal 2000 \\
H$_2$-H   & 1000-2500  & 10000 & Gustafsson, Frommhold 2003 \\
H$_2$-H   & 400-1000  & 6000 & Gustafsson \etal 2003 \\
H-He      & 1500-10000 & 11000 & Gustafsson, Frommhold 2001 \\
\hline\hline
\end{tabular}
\end{table}

The shape of CIE spectra can be found in the literature (see Table
\ref{cie_references} for a list of references), and used for estimating
the cooling rates due to the main CIE processes.
The definition of the monochromatic emission coefficient $j_\nu$
(cfr. Rybicki \& Lightman, eq. 1.15)
\begin{equation}
dE = j_\nu\, dV\, d\Omega\, dt\, d\nu
\end{equation}
(where $E$ is the energy emitted, and $j_\nu$ has units of ${\rm erg\,
cm^{-3}\, ster^{-1}\, s^{-1}\, Hz^{-1}}$) can be easily integrated to
give the luminosity per unit mass $L$
\begin{equation}
L\equiv {{dE}\over{dt\,dm}} = {{dE}\over{dt\,\rho dV}} = {{4\pi}\over{\rho}}
\int{j_\nu d\nu}.
\end{equation}
The gas can be assumed to be in thermal equilibrium, so that
$j_\nu=\alpha_\nu B_\nu(T)$ (where $B_\nu(T)$ is the Planck function
at temperature $T$), and the cooling rate per unit mass due to
emission induced by collisions between particles of species $i$ and
species $j$ in a gas of temperature $T$ and density $\rho$ is
\begin{equation}
L_{i,j}(T,\rho,y_i,y_j)={{4\pi}\over{\rho}} \int
{\alpha_{\nu,i,j}(T,\rho,y_i,y_j) B_\nu(T) d\nu},
\end{equation}
where $y_i$ and $y_j$ are the mass fractions of particles of species $i$
and $j$, respectively. We took the values of the collision-induced absorption
coefficient $\alpha_{\nu,i,j}$ from the references listed in
Table \ref{cie_references}\footnote{These references actually give the
absorption coefficients in ${\rm cm^{-1}\;amagat^{-2}}$, which must be
multiplied by a factor $(\rho y_i/\rho_{0,i})(\rho y_j/\rho_{0,j})$ in
order to find the absorption coefficients appropriate for the given
density and mass fractions. The densities $\rho_{0,i}$ are given by
$\rho_{0,i}=(P_0 m_i)/(k_{\rm B} T_0)$, where $P_0=1\;{\rm atm}$,
$T_0=273.15\;{\rm K}$, $m_i$ is the mass of a particle of species $i$
and $k_{\rm B}$ is the Boltzmann constant.}.

It can be seen that these references provide a complete set of data
only in a relatively narrow temperature range ($1500\;{\rm K} \leq T
\leq 2500\;{\rm K}$). By a fortunate coincidence, this is also the
temperature range where CIE cooling dominates over all the other
cooling mechanisms (see next section), and where CIE effects are more
pronounced. However, we have artificially extended this range to the
full extension of the H$_2$-H$_2$ data range ($400\;{\rm K} \leq T
\leq 7000\;{\rm K}$) by estimating the cooling rates $L_{i,j}(T)$
which are not directly available, through the assumption that the
ratio $L_{i,j}(T)/L_{\rm H_2,H_2}(T)$ is the same as at the closest
known value. This assumption is somewhat arbitrary, but has negligible
effects for a gas where \HH is the dominant species (which is
the most relevant case); in the case of a gas with a high atomic
fraction, and outside the $1500\;{\rm K} - 2500\;{\rm K}$
temperature range, the results shown in Fig. \ref{cie_components} are
only indicative.

\bfig
\psfig{figure=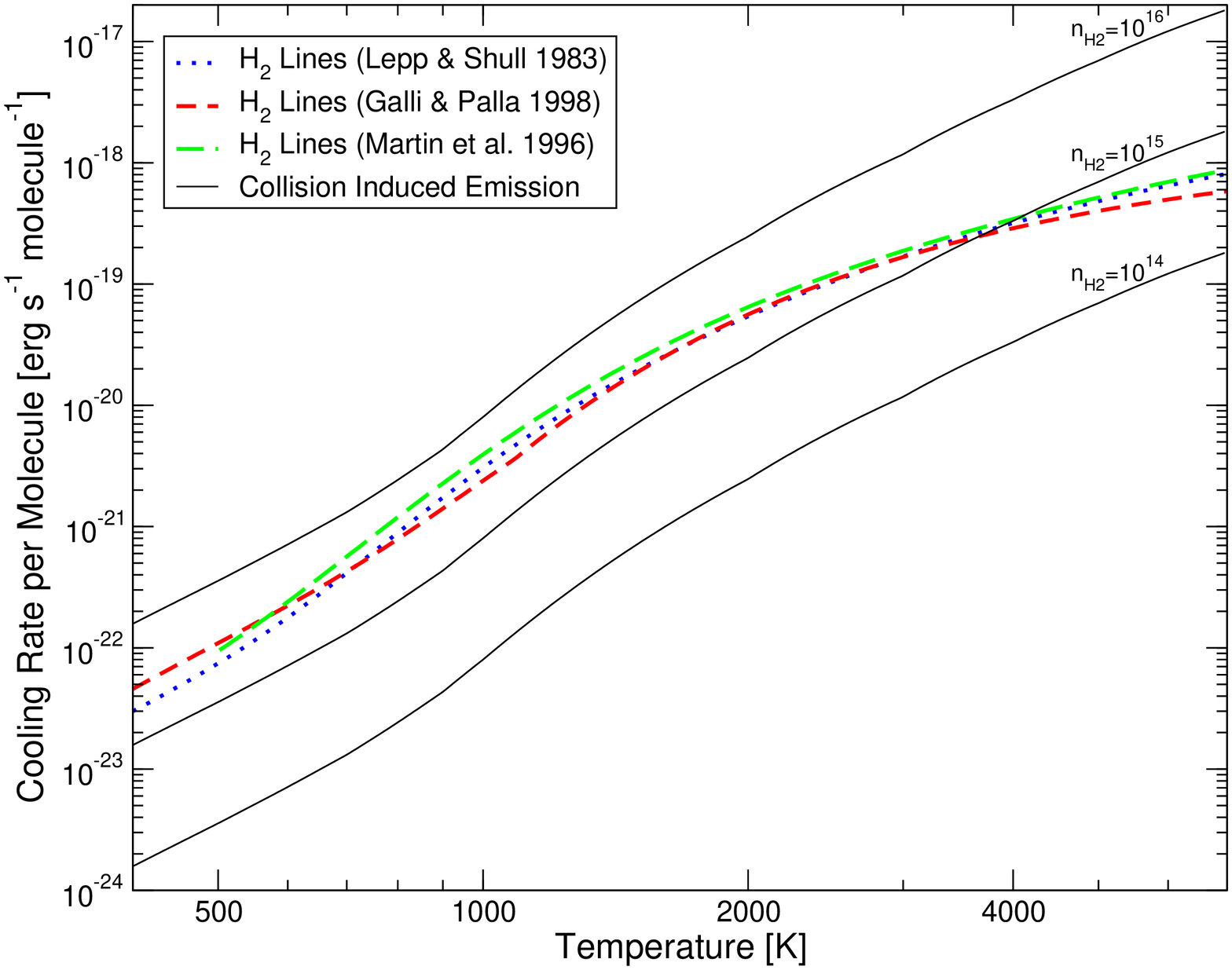,height=10truecm,angle=270}
\caption{Comparison of \HH line cooling rate and H$_2$-H$_2$ CIE cooling
rate. Thick lines denote the line cooling rate as calculated by Galli \&
Palla 1998 (short dashes), Martin, Schwarz \& Mandy 1996 (long dashes)
and Lepp \& Shull 1983 (dots); these were calculated assuming to be in
the high density regime ($n\gsim10^8\;{\rm cm^{-3}}$), where the cooling
per molecule is independent of density. Thin solid lines denote CIE
cooling rates at molecular number densities $n_{\rm H_2}=10^{14}\;{\rm
cm^{-3}}$ (bottom line), $n_{\rm H_2}=10^{15}\;{\rm cm^{-3}}$ (middle
line) and $n_{\rm H_2}=10^{16}\;{\rm cm^{-3}}$ (top line). If the
considered object were optically thin to \HH line cooling, CIE would
become important at number densities $n\gsim 10^{15}\,{\rm cm^{-3}}$.}
\label{cooling_comparison}
\efig

In Fig. \ref{cooling_comparison} we compare the CIE cooling rates of a
pure \HH gas at different densities (similar results hold also for
H$_2$-He and H$_2$-H-He mixtures) with the \HH line cooling rates
calculated by several other authors. CIE cooling overcomes
roto-vibrational line cooling at an \HH number density $n_{\rm H_2}$
between $10^{14}$ and $10^{16}\;{\rm cm^{-3}}$, the exact value
depending on the gas temperature. This values can change because of the
effects of optical depth, and are appropriate only when both CIE cooling
and \HH line cooling occur in an optically thin regime. Instead, both
ON98 and R02 (treating CIE by means of the Planck opacities given by
Lenzuni \etal 1991) found that \HH line cooling starts to be limited by
optical depth at an early stage, when CIE cooling is still optically
thin. As a result, in their models CIE starts to overcome \HH lines at a
relatively low density (corresponding to $T_c\simeq 1600\;{\rm K}$,
$n_{{\rm H_2},c}\simeq 5\times 10^{13}\;{\rm cm^{-3}}$, where the $c$
subscript refers to the conditions in their central regions; see Fig. 4
of R02).

\bfig
\psfig{figure=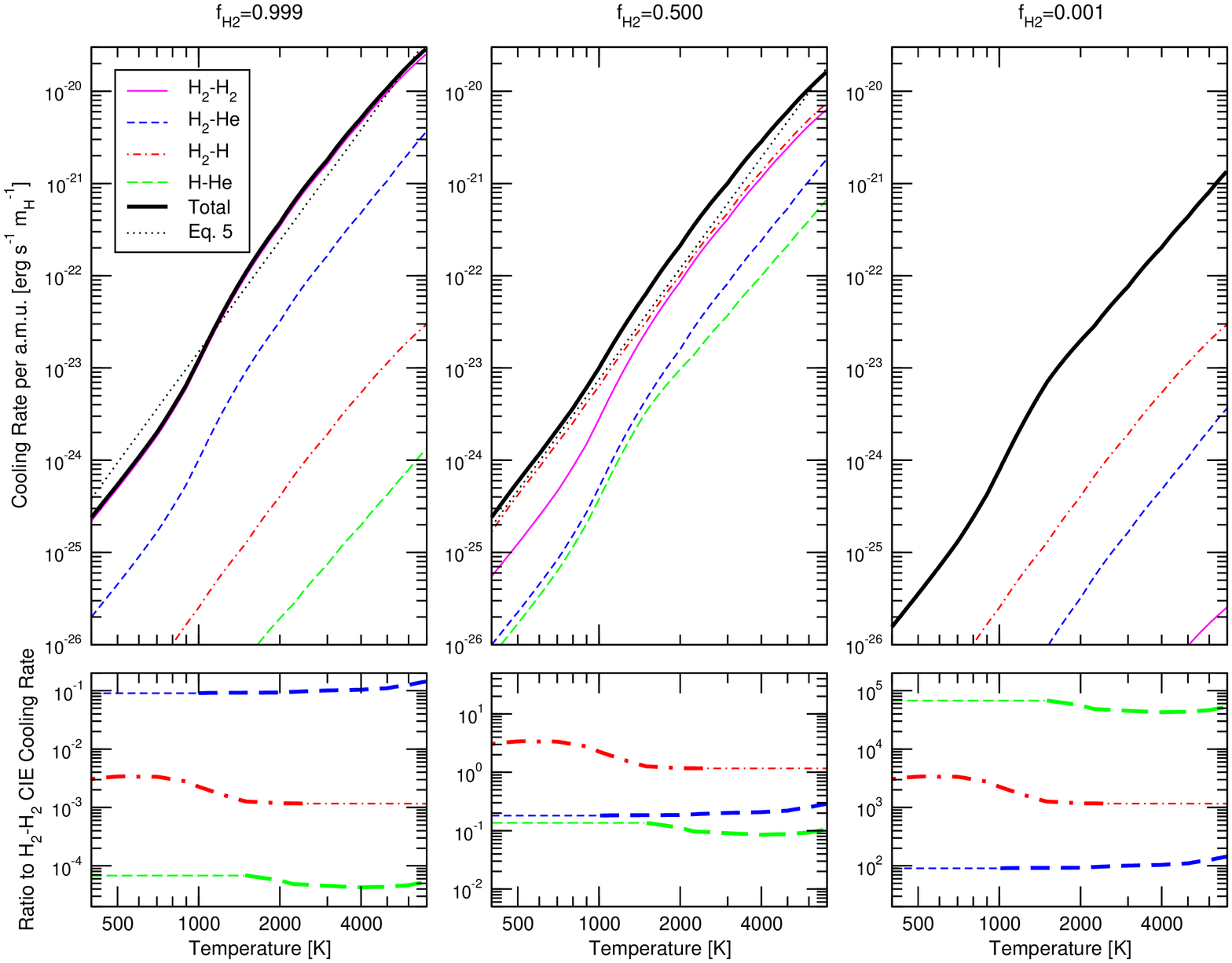,width=17truecm,angle=270}
\caption{Comparison of CIE cooling rates for different kinds of
collisions. Top panels show the cooling rates per unit mass as a
function of temperature, for a gas with different molecular fractions
($f_{\rm H_2}=0.999$, $0.5$ and $0.001$). The thick solid line denotes
the total CIE cooling, the thin lines denote the single components:
H$_2$-H$_2$ (thin solid), H$_2$-He (short dashed), H$_2$-H (dot dashed)
and H-He (long dashed); the dotted lines in the upper left and upper
center panels show the results of eq. (\ref{approximate_cie_cool}). Note
that in the top-left panel the total and H$_2$-H$_2$ lines are
practically indistinguishable, as well as the total and the H-He lines
in the top-right panel. The bottom panels show the ratios of the last
three components to the CIE cooling rate due to H$_2$-H$_2$ collisions;
in these panels the line thickness indicates whether the ratio was taken
from one of the references in table \ref{cie_references} (thick lines)
or it was assumed to be constant at the closest known value (thin
lines). All the quantities were calculated assuming $n=10^{14}\;{\rm
cm^{-3}}$ and $y_{\rm He}=0.25$.}
\label{cie_components}
\efig

The gas chemical composition determines which kind of collisions
contribute most to total cooling. In Fig. \ref{cie_components} we show
the temperature dependency of the cooling rate due to each kind of
collisions for three different molecular fractions, $f_{\rm
H_2}=0.999$, $0.5$ and $0.001$, where $f_{\rm H_2} \equiv {{{2n_{\rm
H_2}}}\over {n_{\rm H} + 2n_{\rm H_2}}}$ is the mass fraction of \HH
(relative to all the H atoms). The He mass fraction is kept
fixed at $y_{\rm He}=0.25$ (the value we will use throughout the rest
of this paper), and the assumed density is
$\rho=1.67\times10^{-10}{\rm g\,cm^{-3}}$ (\ie $n=10^{14}\;{\rm
cm^{-3}}$). As can be expected, \HH collisions (especially
H$_2$-H$_2$) dominate CIE cooling when hydrogen is mostly molecular,
but their importance declines with decreasing \HH fraction, becoming
negligibly small for mostly atomic gas. The total CIE cooling rate for
molecular gas is substantially (20-30 times) larger than for atomic
gas in the same conditions.

\subsubsection{Approximate CIE cooling rate}
In the rest of this paper, the only relevant case will be the one with
$f_{\rm H_2}\simeq 1$. In that case, the total CIE cooling rate (per
unit mass)
\begin{equation}
L_{\rm CIE}(\rho,T,y_{\rm H},y_{\rm H_2},y_{\rm He}) =
L_{\rm H_2,H_2} + L_{\rm H_2,H} +
L_{\rm H_2,He} + L_{\rm H,He}
\end{equation}
can be reasonably approximated by a power law
\begin{equation}
L_{\rm CIE}(\rho,T,y_{\rm H},y_{\rm H_2},y_{\rm He}) \simeq
L_{\rm CIE}(\rho,Y,f_{\rm H_2}) = A_{\rm CIE}\; \rho\, T^\alpha \,
X f_{\rm H_2}
\label{approximate_cie_cool}
\end{equation}
with $\alpha\simeq 4.0$ and $A_{\rm CIE}\simeq 0.072\;{\rm erg\, cm^3\,
g^{-2}\, s^{-1} K^{-4.0}}$, and where $X=1-y_{\rm He}=0.75$ is the total
hydrogen mass fraction.  As can be seen in fig. \ref{cie_components},
this approximation holds down to $f_{\rm H_2}\sim0.5$, although very
roughly.

\subsection{\HH line cooling}

\subsubsection{Optically thin \HH line cooling}
At the moderately high densities we are interested in ($n\gsim10^6\;{\rm
cm^{-3}}$), the populations of \HH roto-vibrational levels can be found
through the LTE assumption, and the \HH line cooling is quite well known
(see Fig. \ref{cooling_comparison}). Here we will use the
LTE rate first given by Hollenbach \& McKee (1979), which was also used
by Galli \& Palla (1998) as an high density limit.

\begin{equation}
L_{\rm lines,HM}(\rho,T,f_{\rm H_2}) =
{{X f_{\rm H_2}}\over{m_H}} {\mathcal{H}}(T_3),
\label{lte_thin_lines_cooling}
\end{equation}
where $L_{\rm lines,HM}$ is the \HH lines cooling rate per unit mass,
$T_3 \equiv {T \over {1000\; {\rm K}}}$, and
\begin{eqnarray}
{\mathcal{H}}(T_3) = &
\left({{9.5\times10^{-22}T_3^{3.76}}\over{1+0.12T_3^{2.1}}}\right)
e^{-\left({{0.13}\over{T_3}}\right)^3} + 
3\times10^{-24}e^{-{{0.51}\over{T_3}}} + \nonumber \\
& 6.7\times10^{-19}e^{-{{5.86}\over{T_3}}} +
1.6\times10^{-18}e^{-{{11.7}\over{T_3}}}\quad {\rm erg\,s^{-1}}.
\label{lte_thin_lines_cooling_ht}
\end{eqnarray}

Eq. (\ref{lte_thin_lines_cooling}) is actually an analytic fit to the
sum of the luminosities of all the roto-vibrational lines (which instead
was used by ON98 and R02):
\begin{equation}
L_{\rm lines,thin}(T) = {{X f_{\rm H_2}}\over{2 m_H}} \sum_{m,n}{h\nu_{m,n}
A_{m,n} \left[{{{2J_n+1}\over{U(T)}} e^{{E_n}\over{k_{\rm B}
T}}}\right]}
\label{lte_sumoflines_cooling}
\end{equation}
where $\nu_{m,n}$ is the frequency of the \HH transition from level $n$
downward to level $m$, $A_{m,n}$ is the spontaneous radiative decay rate
(Einstein coefficient) of the transition, $J_n$ is the rotational
quantum number associated with level $n$, $E_n$ is the energy of level
$n$ and $U(T)=\sum_i {(2J_i+1)e^{E_i/(k_{\rm B} T)}}$ is the \HH
partition function at temperature $T$.  The sum in
(\ref{lte_sumoflines_cooling}) is over all \HH lines, with each term
representing the total luminosity of an \HH molecule in the considered
line. We note that the term in square brackets represents the fraction
of \HH molecules in the roto-vibrational level $n$.

\subsubsection{Optically thick \HH line cooling}
The above cooling rates (eq. \ref{lte_thin_lines_cooling} and
\ref{lte_sumoflines_cooling}) can only be applied to optically thin
objects. In reality, as 3-body reactions start transforming the bulk
of hydrogen into molecular form, the core regions of the contracting
protostellar clouds quickly become optically thick to \HH line
radiation (ON98, R02). As a result, the effective cooling rate falls
well below the one predicted in the optically thin case.

It is possible to find a reasonable approximation to the ``effective''
cooling rate in the central regions of a collapsing protostellar
object. The density profile of such an object can be approximated with
a central flat ``core'' (where also temperature and chemical
composition are approximately constant) surrounded by an ``envelope''
where the density decreases as a power-law ($\rho \propto r^{-2.2}$,
according to both ON98 and R02). Such a density profile is predicted
by the Larson-Penston self-similar solution (Larson 1969; Penston
1969), which applies fairly well. Another correct prediction that can
be inferred from the Larson-Penston solution is that the mass of the
central ``core'' is of the order of the Bonnor-Ebert mass (\ie the
critical mass for gravitational collapse of an isothermal sphere given
an external pressure $P_{\rm ext}$; see Ebert 1955 and Bonnor 1956),
as calculated using the central value of temperature ($T_c$), number
density ($n_c=\rho_c/m_{\rm H}$), molecular weight ($\mu_c$) and mean
adiabatic index ($\gamma_c$)
\begin{equation}
M_{{\rm BE},c} \simeq 20\;{\rm M_\odot}
\left({T_c\over{\rm 1\;K}}\right)^{3/2}
\left({n_c\over{\rm 1\;cm^{-3}}}\right)^{-1/2} \mu_c^{-2} \gamma_c^2;
\label{bonnor_ebert_mass}
\end{equation}
from this a ``Bonnor-Ebert radius'' can be estimated
\begin{equation}
R_{{\rm BE},c} =
\left({{{3M_{{\rm BE},c}}\over {4 \pi n_c m_H}}}\right)^{1/3} \simeq
1.8\times10^{19}\;{\rm cm}\;
\left({T_c\over{\rm 1\;K}}\right)^{1/2}
\left({n_c\over{\rm 1\;cm^{-3}}}\right)^{-1/2} \mu_c^{-2/3} \gamma_c^{2/3}.
\label{bonnor_ebert_radius}
\end{equation}
The \HH column density from the centre to infinity is then 
\begin{eqnarray}
\label{nh2_column}
N_{{\rm H_2},c} & \simeq &
R_{{\rm BE},c} {{n_c X f_{{\rm H_2},c}} \over 2} +
\int_{R_{{\rm BE},c}}^{\infty}{ {{n_c X f_{\rm H_2}(r)} \over 2}
\left({r\over{R_{{\rm BE},c}}}\right)^{-2.2} dr} = \nonumber \\
& = & R_{{\rm BE},c} {{n_c X f_{{\rm H_2},c}}\over 2}
\left\{{1+\int_1^\infty{[f_{\rm H_2}(x R_{{\rm BE},c})/f_{{\rm H_2},c}]
x^{-2.2}dx}}\right\}
\end{eqnarray}
where $x\equiv r/R_{{\rm BE},c}$.
This last equation leads us to define a parameter $\xi$ such that
\begin{equation}
N_{{\rm H_2},c} \equiv \xi R_{{\rm BE},c} {{n_c X f_{{\rm H_2},c}}\over 2}
\end{equation}
and whose approximate value (cfr. eq. \ref{nh2_column}) is
\begin{equation}
\xi \simeq
1+\int_1^\infty{[f_{\rm H_2}(x R_{{\rm BE},c})/f_{{\rm H_2},c}]x^{-2.2}dx}.
\label{xi_approx}
\end{equation}

Since we are considering objects in which $f_{\rm H_2}$ is maximum at
the centre, equation (\ref{xi_approx}) implies that $1\leq\xi\leq1.8$,
but in Table \ref{bonnor_ebert_comparison_table} (see columns 6 and 7)
we show that this is not the case. The main reasons are that the density
inside the core actually decreases from the centre outward, and that the
Bonnor-Ebert mass provides an estimate of the core size which is not
very accurate (Table \ref{bonnor_ebert_comparison_table}, columns 4 and
5). In the following we will just set $\xi$ to values in agreement with
the results of R02 (for example, Fig. \ref{rho_thick_cool} was obtained
using $\xi=0.20$).

\begin{table}
\caption{Comparison of the core properties as found at several stages of
R02 calculations, to the values employed in the text: mass and radius are
compared to the Bonnor Ebert values (eq. \ref{bonnor_ebert_mass}\ and
\ref{bonnor_ebert_radius}), while the column densities ($N_{\rm
H_2,\infty}$ and $N_{\rm H_2,core}$, representing the column density from
the centre to infinity and from the centre to the edge of the core,
respectively) are normalized to
${{R_{{\rm BE},c} n_c X f_{{\rm H_2},c}}/2}$. In order to extract
core data from the simulations, we defined it as the central region
where the density dependence on radius is flatter than $\rho\propto r^{-1}$.}
\begin{tabular}{ccccccc}
\hline\hline $T_c[{\rm K}]$ & $n_c[{\rm a.m.u.\,cm^{-3}}]$ & $f_{{\rm
H_2},c}$ & $M_{\rm core}/M_{{\rm BE},c}$ & $R_{\rm core}/R_{{\rm BE},c}$
& ${{N_{{\rm H2},\infty}}\over{R_{{\rm BE},c} n_c X f_{{\rm
H_2},c}/2}}=\xi_\infty$ & ${{N_{\rm H_2,core}}\over{R_{{\rm BE},c} n_c X
f_{{\rm H_2},c}/2}}=\xi_{\rm core}$\\ \hline 300 & $9.1\times10^6$ &
0.001 & 0.24 & 0.69 & 0.62 & 0.41\\ 401 & $7.2\times10^7$ & 0.002 & 0.22
& 0.67 & 0.54 & 0.41\\ 502 & $1.0\times10^9$ & 0.02 & 0.20 & 0.64 & 0.45
& 0.39\\ 594 & $5.5\times10^9$ & 0.11 & 0.22 & 0.67 & 0.48 & 0.42\\ 740
& $2.8\times10^{10}$ & 0.39 & 0.29 & 0.76 & 0.60 & 0.51\\ 913 &
$1.4\times10^{11}$ & 0.72 & 0.78 & 1.17 & 0.81 & 0.62\\ 1328 &
$5.5\times10^{12}$ & 0.98 & 0.49 & 0.88 & 1.23 & 0.72\\ 1636 &
$4.8\times10^{13}$ & 0.98 & 0.57 & 0.95 & 1.31 & 0.73\\ 1799 &
$1.8\times10^{14}$ & 0.96 & 0.58 & 0.94 & 1.40 & 0.79\\ 2007 &
$1.1\times10^{15}$ & 0.92 & 0.64 & 0.99 & 1.37 & 0.74\\ 2200 &
$8.0\times10^{15}$ & 0.91 & 0.69 & 1.00 & 1.43 & 0.77\\ 2337 &
$4.6\times10^{16}$ & 0.92 & 0.85 & 1.12 & 1.41 & 0.77\\ \hline\hline
\end{tabular}
\label{bonnor_ebert_comparison_table}
\end{table}

Inside the ``core'', where we can assume that the temperature is
reasonably constant, the cross section due to a transition from level
$m$ to level $n$ is (see Lang 1980, eq. 2-69)
\begin{equation}
\sigma_{m,n}(\nu) = \phi(T_c,\nu,\nu_{m,n})
{{c^2}\over{8\pi\nu_{m,n}^2}}
\left({e^{{h\nu_{m,n}}\over{k_{\rm B} T_c}}-1}\right) 
A_{m,n},
\end{equation}
where $\phi$ is the line profile. This cross section applies only to \HH
molecules in the $m$ roto-vibrational level, that is to a fraction
${{2J_m+1}\over{U(T_c)}}e^{{E_m}\over{k_{\rm B} T_c}}$ of all the \HH
molecules.

In the range of conditions we are considering, the line profile is
determined by Doppler broadening, with typical width
\begin{equation}
\Delta\nu_{\rm D}(T_c) = {{\nu_{m,n}}\over c} \sqrt{{2 k_{\rm B}
T_c}\over{m_{\rm H_2}}}.
\end{equation}
Here, we will assume that the line profile can be simplified to
\begin{equation}
\phi(T_c,\nu,\nu_{m,n}) =
      	\left\{{
	\begin{array}{ll} 
	{1\over{2\Delta\nu_{\rm D}(T_c)}} &
	\ {\rm if}\ |\nu-\nu_{m,n}|\le\Delta\nu_{\rm D}(T_c)\quad {\rm
      	(``inside''\ the\ line)};\\
	0 & \ {\rm if}\ |\nu-\nu_{m,n}|\le\Delta\nu_{\rm D}(T_c)\quad {\rm
      	(``outside''\ the\ line)},
        \end{array}}\right .
\end{equation}
and we will neglect the Doppler shifts due to the bulk
motions\footnote{At the considered stages, the collapsing cloud is not
in hydrostatic equilibrium, not even in the core: bulk velocities can
amount to a few $\times 10^5\;{\rm cm\,s^{-1}}$. The resulting frequency
shift is smaller than the line width, although not completely
negligible.}.  With this approximation, and averaging over all \HH
molecules, the mean cross section of a generic \HH molecule to a
photon with frequency within $\Delta\nu_{\rm D}(T_c)$ from $\nu_{m,n}$
is
\begin{equation}
\sigma_{m,n} = {{c^3}\over{16\pi\nu_{m,n}^3}}
\sqrt{{m_{\rm H}}\over{k_{\rm B} T_c}}
\left({e^{{h\nu_{m,n}}\over{k_{\rm B} T_c}}-1}\right) A_{m,n}
{{2J_m+1}\over{U(T_c)}} e^{{E_m}\over{k_{\rm B} T_c}}
\end{equation}
and the resulting optical depth from centre to infinity is then
\begin{equation}
\tau_{m,n} \simeq 0.53\; \xi
\left[{
\left({e^{{h\nu_{m,n}}\over{k_{\rm B} T_c}}-1}\right)
{{2J_m+1}\over{U(T_c)}} e^{{E_m}\over{k_{\rm B} T_c}}
}\right]
Xf_{{\rm H_2},c}
\left({n_c\over{\rm 1\;cm^{-3}}}\right)^{1/2}
\left({{\nu_{m,n}}\over{10^{12}\;{\rm Hz}}}\right)^{-3}
\left({{A_{m,n}}\over{10^{-9}\;{\rm s^{-1}}}}\right)
\left({{\gamma_c}\over{\mu_c}}\right)^{2/3}.
\label{approx_thick_h2_tau}
\end{equation}

\bfig
\psfig{figure=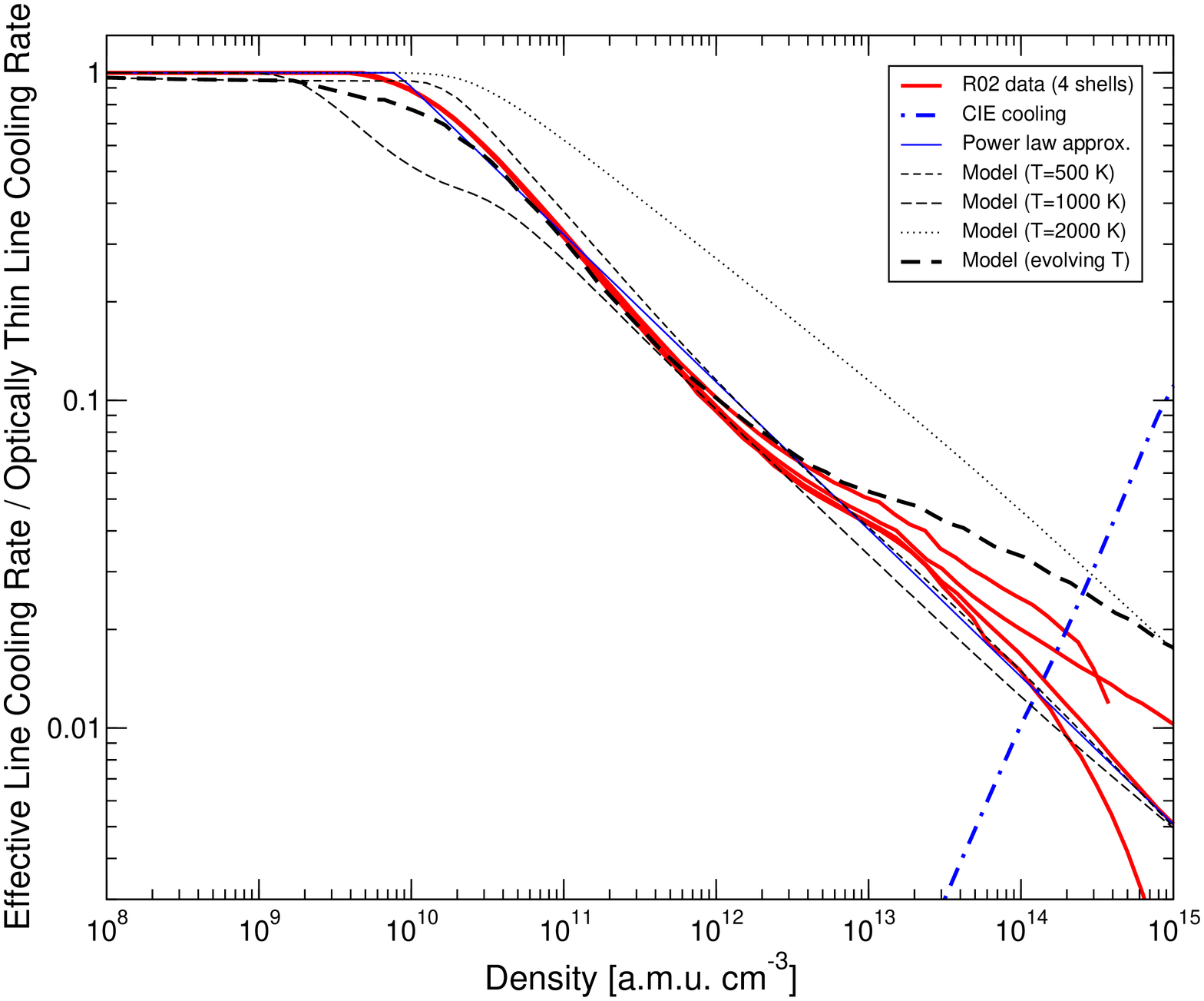,height=10truecm,angle=270}
\caption{Ratio of cooling rates to the optically thin \HH line cooling
rate, as a function of density. The solid thick lines show the evolution
of the effective (that is, as influenced by optical depth) \HH line
cooling rate as calculated by R02 in four different shells (the lowest
one being the central shell), and the thick dot-dashed line shows the
CIE cooling rate. The thin dashed, long dashed and dotted lines show the
predictions of eq. (\ref{h2_approx_thick_cooling}) when assuming
$T_c=500\;{\rm K}$, $1000\;{\rm K}$ and $2000\;{\rm K}$, respectively
(note that these lines are not in monotonic order; the $T_c=500\;{\rm
K}$ line actually lies between the other two);
instead, the thick dashed line shows the prediction of the same equation
when the temperature evolution of the central shell is kept into
account. Finally, the thin solid line shows the simple power law
approximation of eq. (\ref{powerlaw_thick_cool}). All the curves making
use of eq. (\ref{h2_approx_thick_cooling}) assume $\xi=0.20$.}
\label{rho_thick_cool}
\efig

From this result, we can obtain an average cooling rate (per unit mass)
inside the ``core'' by estimating the total luminosity of the core as
seen from outside its surface. This can be done by assuming that the
``core photosphere'' (the region of the core where the optical depth to
the surface is $\leq 1$) radiates in the optically thin limit,
while the interior (where the optical depth to the surface is $> 1$)
does not contribute to the total luminosity.

Since the optical depths of the various lines are different, this
amounts introducing correcting factors the for cooling rate of each line 
into eq. (\ref{lte_sumoflines_cooling}):
\begin{equation}
L_{\rm lines,thick}(T) = {{X f_{\rm H_2}}\over{2 m_H}}
\sum_{m,n}{h\nu_{m,n} A_{m,n} \left[{{{2J_n+1}\over{U(T)}}
e^{{E_n}\over{k_{\rm B} T}}}\right]}\; V_{m,n}\; G_{m,n},
\label{h2_approx_thick_cooling}
\end{equation}
where $V_{m,n}$ is the ratio between the volume of the ``core
photosphere'' for the considered $(m,n)$ transition and the total core
volume, given by
\begin{equation}
V_{m,n} =
      	\left\{{
	\begin{array}{ll} 
	1 &
	\ {\rm if}\ \tau_{m,n} \leq 1\\
	\left[{1-\left({1-{1\over{\tau_{m,n}}}}\right)^3}\right] &
	\ {\rm if}\ \tau_{m,n} > 1.
        \end{array}}\right .
\label{volume_correction_factor}
\end{equation}
while the geometrical correction
\begin{equation}
G_{m,n} =
      	\left\{{
	\begin{array}{ll} 
	1 &
	\ {\rm if}\ \tau_{m,n} \leq 1\\
	1-{1\over2}\left[{{2\over{\tau_{m,n}}}
	\left({1-{1\over{2\tau_{m,n}}}}\right)}\right]^{1/2} &
	\ {\rm if}\ \tau_{m,n} > 1.
        \end{array}}\right .
\label{geometrical_correction_factor}
\end{equation}
accounts for the radiation emitted towards the interior.

We note that both the form of $V_{m,n}$ as the ratio of
the photospheric to the core volume, and the photospheric volume itself
(at least in the optically thick case) clearly depend on the assumption
that inside the core density and temperature are constant.

\bfig
\psfig{figure=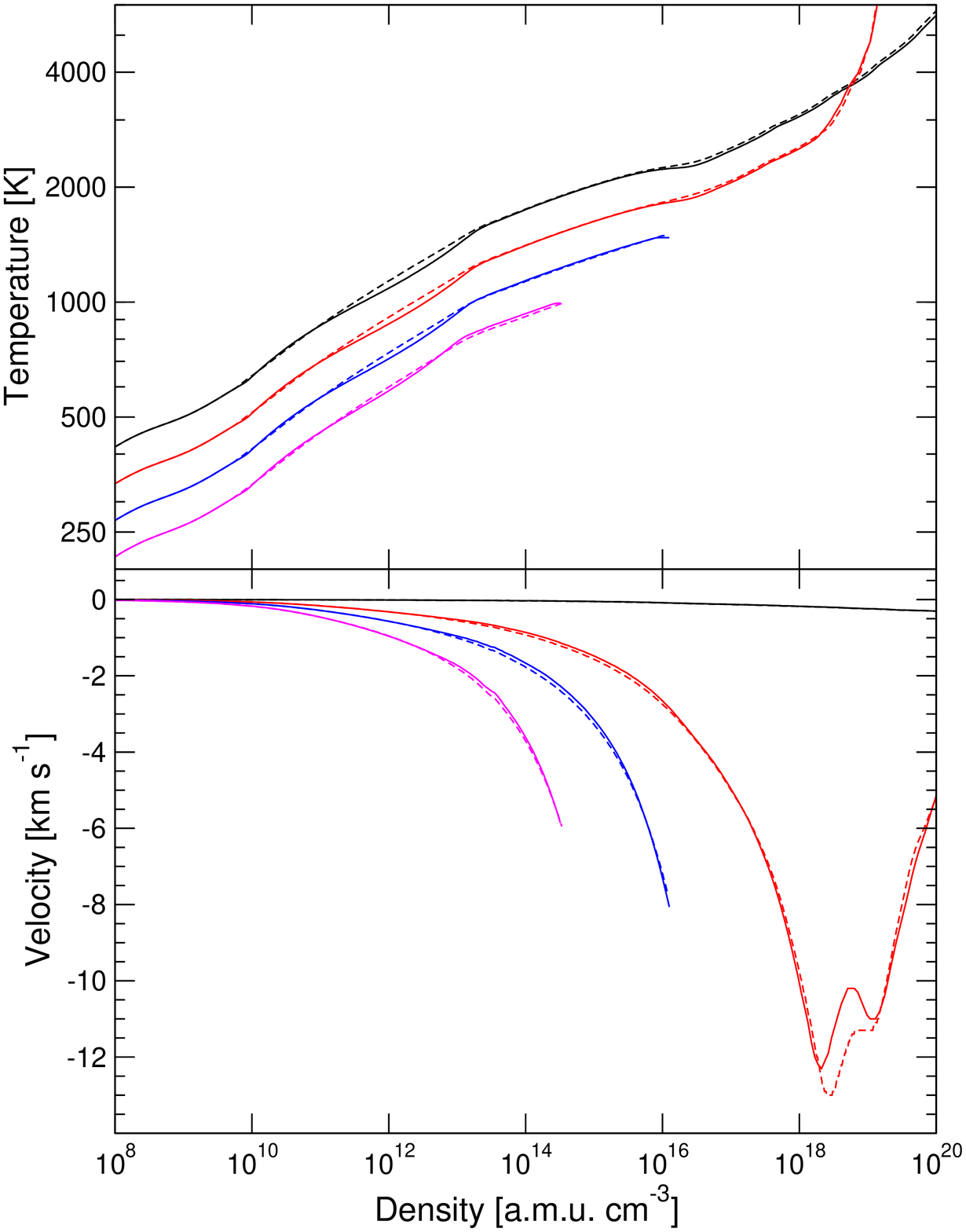,height=12truecm}
\caption{Comparison of the results of a modified version of the R02 code
(solid lines) and of the original one (dashed lines).  In the modified
version the cooling rate is given by eq. (\ref{r02path_l_function}),
which is the sum of the approximate cooling rates discussed in Section
2.2.2 (\HH lines) and in Sections 2.2.1 and 3.3 (CIE); instead, in the
original code the cooling rate is estimated through radiative transfer
calculations. In the top panel we compare the evolution of four shells
(including the central one) in the $n-T$ plane; since all the lines were
very close to each other, we artificially shifted down the results of
the three non-central shells (with the bottom lines corresponding to the
outermost shell). In the bottom panel we compare the evolution of the
same four shells in the $n-v$ plane. Note that two of our zones do not
reach the maximum density shown in these plots, so the corresponding
lines stop at $n\simeq10^{14}\;{\rm cm^{-3}}$ and $n\simeq10^{16}\;{\rm
cm^{-3}}$}
\label{evolution_comparison}
\efig

In Fig. \ref{rho_thick_cool} we compare the results of
eq. (\ref{h2_approx_thick_cooling}) to the findings of R02. Provided
that a sensible value of $\xi$ is chosen (in Fig. \ref{rho_thick_cool}
we are using $\xi=0.20$), the agreement is very good up to number
densities of $n\sim 10^{13}\;{\rm cm^{-3}}$, and remains reasonable
(within a factor of 2) up to the density where CIE cooling starts to be
dominant ($n\sim 10^{14}\;{\rm cm^{-3}}$). The differences can be
partially explained by noting that at the highest densities the $\xi$ values
reported in Table \ref{bonnor_ebert_comparison_table} are substantially
higher than the adopted value of 0.20. It is also
remarkable that the agreement remains good also when we consider the R02
results for non-central shells, provided that their own physical
properties ($T,\ n\ {\rm and}\ f_{\rm H_2}$) are plugged into
eq. (\ref{approx_thick_h2_tau}) instead of the central ones.

In the next sections we need an analytic expression for the cooling
rate which is less cumbersome than
eq. (\ref{h2_approx_thick_cooling}). We choose to employ a very simple
approximation of the form
\begin{equation}
L_{\rm lines,thick}(T) =
L_{\rm lines,thin}(T) \min{(1,(n/n_0)^{-\beta})}
\label{powerlaw_thick_cool}
\end{equation}
with $n_0=8\times10^{9}\;{\rm cm^{-3}}$ (equivalent to
$\rho_0=1.34\times 10^{-14}\;{\rm g\,cm^{-3}}$) and $\beta=0.45$. 

\bfig
\psfig{figure=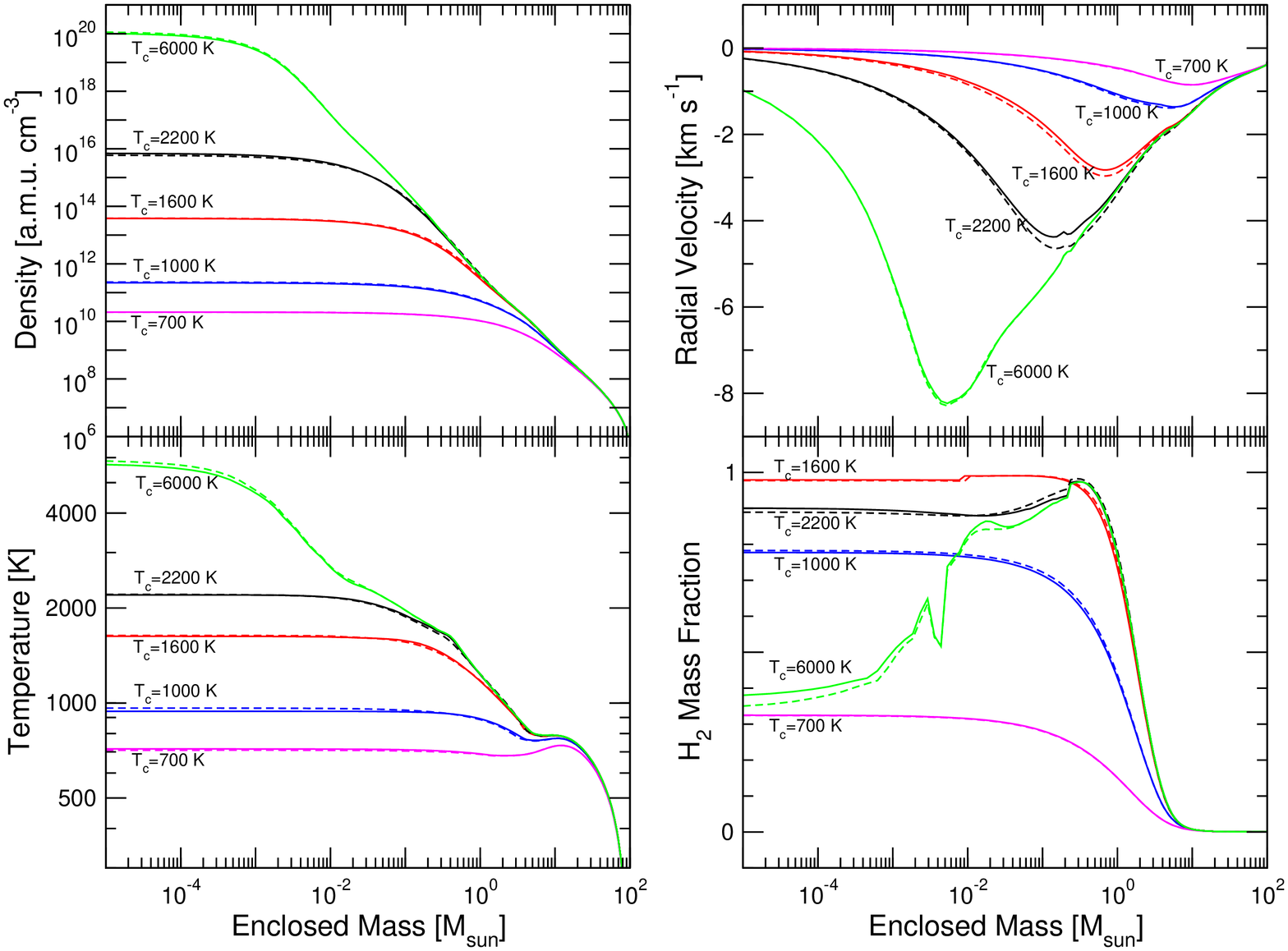,height=12truecm,angle=270}
\caption{Comparison of the results of the original and of the modified
R02 code (see caption of fig. \ref{evolution_comparison}). The various
panels show the spatial profiles of density (top left), velocity (top
right), temperature (bottom left) and molecular fraction (bottom right)
at five different stages (specified through their central temperature)
when the simplification in the \HH lines cooling rate is
relevant. Dashed lines show the results of the original code, while
solid lines show the result of the simplified one.  The spatial
coordinate is specified through the Lagrangian coordinate of R02, {\it
i.e.}\ the enclosed mass.}
\label{structure_comparison}
\efig

We have tested the validity of both equations
(\ref{h2_approx_thick_cooling}) and (\ref{powerlaw_thick_cool}) by
substituting it to the radiative transfer treatment inside the code used
by R02 for their simulations, and comparing the results with those of
the original code; in these tests we also added an approximate CIE
contribution as can be found in eq. (\ref{r02path_l_function}). As can
be seen in Fig. \ref{evolution_comparison} and
\ref{structure_comparison} (which show the results obtained using
eq. \ref{powerlaw_thick_cool}), the evolution is strikingly similar, not
only in the central zones but also in the outer regions.  So, we have
been able to predict the {\it global} effect of radiative transfer on
the \HH line cooling with a simple model based on purely {\it local}
properties. It is quite remarkable (and partly unexpected) that a
treatment which was believed to apply just to the centre of the ``core''
can be applied also to the non-central regions.

The results of eq. (\ref{h2_approx_thick_cooling}) and
(\ref{powerlaw_thick_cool}) could be used for extending full 3-D
simulations to densities up to $n\sim 10^{16}\;{\rm cm^{-3}}$. However,
we have to remark that such an approach is likely to be correct only if
the collapsing object remains approximately spherical at all stages, and
if the spatial profiles (of density, temperature, chemical composition
etc.) actually resemble the 1-D results, at least until the moment when
CIE cooling starts to overcome \HH lines ($n\gsim10^{14}\;{\rm
cm^{-3}}$).

\section{Instabilities}

In this section, we investigate whether a collapsing protostellar cloud
can become unstable to fragmentation by means of two different
instability criteria.

\subsection{Timescales comparison}\label{timescales_comparison}

A classical criterion for fragmentation is that the cooling time $t_{\rm
cool}$ must be shorter that the dynamical timescale

\begin{equation}
t_{\rm dyn}\equiv \left({{3\pi}\over{16G\rho}}\right)^{1/2}.
\label{tdyn}
\end{equation}

In Fig. \ref{time_comp}, we show the evolution of the ratios of three
relevant timescales to the dynamical timescale, as determined from R02
data about the central evolution of a proto-stellar object (in the rest
of this paper, we will refer to this evolutionary track as the ``R02
path'').

These timescales are:\\
- the radiative cooling timescale (accounting for the effects of cooling)
\begin{equation}
t_{\rm rad}= {{[1/(\mu m_{\rm H})]k_{\rm B}T}\over{L_{\rm rad}(\gamma-1)}}
\label{trad}
\end{equation}
where $L_{\rm rad}$ is the total radiative cooling rate per unit mass
($L_{\rm rad}=L_{\rm lines}+L_{\rm CIE}$ in the conditions we are
considering) and $\gamma$ is the mean adiabatic index of the gas
(cfr. ON98);\\
- the thermo-chemical timescale (accounting for the thermal effects of
chemical reactions, namely of formation or disruption of \HH)
\begin{equation}
t_{\rm chem} =
{{(n/\mu)k_{\rm B}T}\over{|\dot{E}_{\rm chem}|(\gamma-1)}} \simeq
{{(n/\mu)k_{\rm B}T}\over{|\dot{n}_{\rm H_2}|\chi_{\rm H_2}(\gamma-1)}},
\label{tchem}
\end{equation}
where $\dot{E}_{\rm chem}\simeq-\dot{n}_{\rm H_2}\chi_{\rm H_2}$ is the
rate of variation of chemical energy (per unit volume), $\dot{n}_{\rm
H_2}$ is the rate of \HH formation (per unit volume) and $\chi_{\rm
H_2}=4.48\;{\rm eV}$ is the binding energy of \HH;\\
- the ``effective'' cooling timescale (accounting for
both chemical reactions and radiation):
\begin{equation}
t_{\rm cool} =
{{[1/(\mu m_{\rm H})]k_{\rm B}T} \over
{(L_{\rm rad} - \dot{E}_{\rm chem}/\rho)(\gamma-1)}}.
\label{tcool}
\end{equation}


\bfig
\psfig{figure=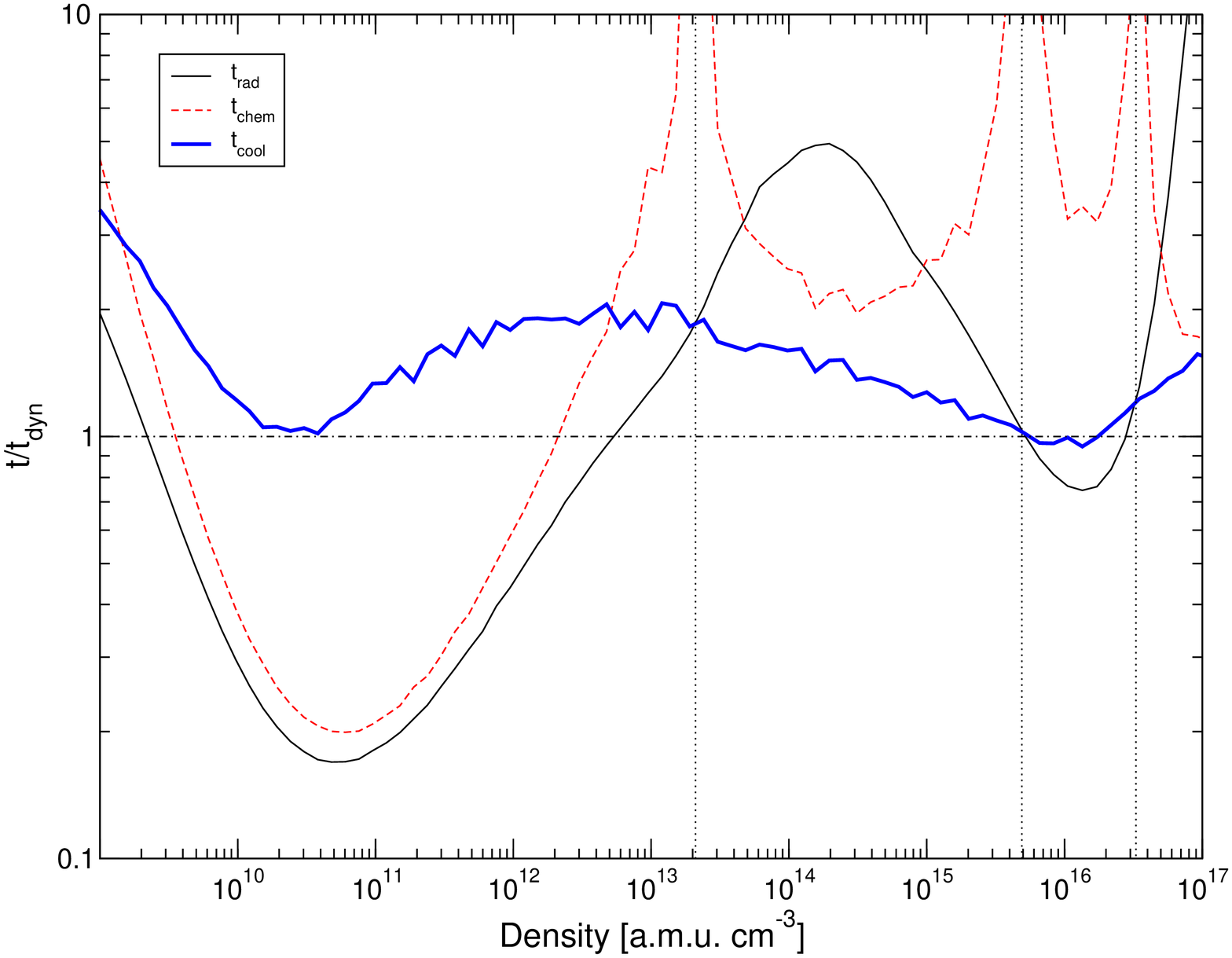,height=10truecm,angle=270}
\caption{Evolution of the ratio of the radiative cooling timescale
$t_{\rm rad}$ (thin solid line), chemical timescale $t_{\rm chem}$
(dashed line) and cooling timescale $t_{\rm cool}$ (thick solid line) to
the dynamical timescale $t_{\rm dyn}$, as a function of density. The
data come from the protostellar collapse simulations of R02 and refer to
the central zone (``R02 path''). The three vertical dotted lines at
densities $\sim 3\times 10^{13}$, $\sim 5\times 10^{15}$ and $\sim
3\times 10^{16}\;{\rm m_H\,cm^{-3}}$ show where the chemical heating
term vanishes (implying $t_{\rm chem}=\infty$ and $t_{\rm cool}=t_{\rm
rad}$) and changes sign. The horizontal dot-dashed line just visualizes
the criterion $t<t_{\rm dyn}$.}
\label{time_comp}
\efig

It can be seen in Fig. \ref{time_comp} that the ratio $t_{\rm
cool}/t_{\rm dyn}$ is close to 1 for most of the considered evolution,
although it never really crosses this critical value. Since the above
criterion is only a sufficient condition for instability,  we
consider a more detailed instability criterion next.

\subsection{CIE induced instabilities}

\subsubsection{Chemo-thermal instability criterion}
Silk (1983; see also Sabano \& Yoshii 1977) considered a gas cloud in
chemical and thermal equilibrium and investigates the conditions for
the growth of isobaric perturbations.  Assuming infinitesimal
perturbations of the form $T=T_1\exp(ikx+\omega t)$, a dispersion
relation
\begin{equation}
A\omega^2 + B\omega + C = 0
\label{dispersion_relation}
\end{equation}
is obtained, which can be used to assess the stability of a given gas
cloud. Similar results were obtained by Saio \& Yoshii (1986), and OY03,
who were able to drop the assumption of chemical and thermal equilibrium.

We use the values of the coefficients $A$, $B$ and $C$ which are given
by equations (A19-A27) of OY03: they depend on the temperature $T$, on
the density $\rho$, on the molecular mass fraction $f_{\rm H_2}$, on the
two functions ${\mathcal{F}}(\rho,T,f_{\rm H_2})\equiv {{df_{\rm
H_2}}\over{dt}}$ (rate of change of the molecular mass fraction) and
${\mathcal{L}}(\rho,T,f_{\rm H_2})$ (cooling rate per unit mass), and on
their partial derivatives ${\mathcal{F}}_\rho = \partial
{\mathcal{F}}/\partial \rho$, ${\mathcal{F}}_f = \partial {\mathcal{F}}
/ \partial f_{\rm H_2}$, ${\mathcal{F}}_T = \partial {\mathcal{F}} /
\partial T$, ${\mathcal{L}}_\rho = \partial {\mathcal{L}} /\partial
\rho$, ${\mathcal{L}}_f = \partial {\mathcal{L}} / \partial f_{\rm
H_2}$, and ${\mathcal{L}}_T = \partial {\mathcal{L}} / \partial T$.

Since $A$ must be positive, it is possible to have growing perturbations
(\ie, ${\rm Re}(\omega)>0$) if and only if $C<0$, with (cfr. OY03,
appendix A)
\begin{eqnarray}
C=&
-{{\mu m_{\rm H}}\over{k_{\rm B} T}}
(T{\mathcal L}_T - \rho{\mathcal L}_\rho - {\mathcal L})
({\mathcal F}_f + {\mu\over 2}\rho{\mathcal F}_\rho +
                  {{\mathcal F}\over{2-f_{\rm H_2}}}) \nonumber\\
&+{{\mu m_{\rm H}}\over{k_{\rm B} T}}
({\mathcal L}_f + {\mu\over 2}\rho{\mathcal L}_\rho +
                  {1\over{6-f_{\rm H_2}}}{\mathcal L})
(T{\mathcal F}_T - \rho{\mathcal F}_\rho) \nonumber\\
&+{\mu\over 2} {\mathcal F}
\left[{ {\mu\over{3-f_{\rm H_2}/2}}({1\over2}+{\chi\over{k_{\rm B}T}})
        (T{\mathcal F}_T - \rho{\mathcal F}_\rho)
     -{\chi\over{k_{\rm B}T}}({\mathcal F}_f +
         {\mu\over 2}\rho{\mathcal F}_\rho +
         {{\mathcal F}\over{2-f_{\rm H_2}}})
}\right] \nonumber\\
&+{\mu\over{(3-f_{\rm H_2}/2)t_{\rm ff}}}
(T{\mathcal F}_T - \rho{\mathcal F}_\rho),
\label{instability_criterion}
\end{eqnarray}
where $\mu = (1-f_{\rm H_2}/2)^{-1}$ is the mean molecular weight (OY03
assumed a pure H-\HH gas) and $t_{\rm ff}=[3\pi/(32G\rho)]^{1/2}$ is the
free fall timescale.

\subsubsection{The \HH formation and cooling rates (${\mathcal{F}}$ and
${\mathcal{L}}$)}

We now apply the above instability criterion
(eq. \ref{instability_criterion}) to the density and temperature regime
where optically thin CIE cooling could be dominant. For this reason, we
will only consider the density range $10^{13}\;{\rm cm^{-3}} \le n_p \le
10^{17}\;{\rm cm^{-3}}$ (where $n_p=\rho X / m_H = n_H + 2n_{\rm H_2}$
is the number density of protons); we will also restrict ourselves to
the temperature range $400\;{\rm K} \le T \le 3000\;{\rm K}$,
because of the limitations of the approximate cooling rate
we are going to assume (eq. \ref{approximate_cie_cool}).

In this region of the density-temperature phase space, \HH formation and
dissociation proceeds through the three-body reactions
described by PSS83, so that the function ${\mathcal{F}}$ can be written as
\begin{equation}
{\mathcal{F}} \equiv {{df_{\rm H_2}}\over{dt}} =
n_f [2nk_4(1-f_{\rm H_2})^2-k_5f_{\rm H_2}]
\label{molecular_formation_rate}
\end{equation}
with
\begin{equation}
n_f = n_p \left({1-{{15f_{\rm H_2}}\over{16}}}\right),
\end{equation}
while $k_4$ and $k_5$ (the notation comes from PSS83) are the reaction rates
for the formation and disruption of \HH, respectively. We use the PSS83
formation rate ($k_4$), but we modify the dissociation rate ($k_5$) in
order to better approximate the ``high density'' ($n\gsim 10^9\;{\rm
cm^{-3}}$) results of Martin, Schwarz \& Mandy (1996):
\begin{eqnarray}
k_4 = & 5.5\times 10^{-29}\, T^{-1}\quad{\rm cm^{6}\,s^{-1}}\\
k_5 = & 2.2\times 10^{-9}\, T^{0.2}\, e^{-{{B_5}\over T}}\,
(1-e^{-{{C_5}\over T}})\quad{\rm cm^{3}\,s^{-1}}
\end{eqnarray}
where $B_5=51800\;{\rm K}$, $C_5=6000\;{\rm K}$.

At these high densities ON98 and R02 showed that we can safely assume
that chemical equilibrium is attained, so that the
condition ${\mathcal{F}}=0$ can be coupled with equation
(\ref{molecular_formation_rate}) in order to get the equilibrium \HH
fraction $f_0(\rho,T)$:
\begin{equation}
f_0(\rho,T)=1+{{k_5}\over{4 n_p k_4}}
\left[{1-\left({1+{{8 n_p k_4}\over{k_5}}}\right)^{1/2}}\right].
\label{equilibrium_fh2}
\end{equation}
However, we note that the third addendum in
eq. (\ref{instability_criterion}) introduces a strong dependence of the
value of $C$ upon the actual \HH abundance, and that even relatively
small differences (a few percent) in $f_{\rm H_2}$ can lead to
significant discrepancies in the results, as we will show in the next
subsection.

With regard to the ``luminosity'' ${\mathcal{L}}$, in this subsection we
will neglect all forms of radiative cooling except CIE optically
thin cooling. In addition, the internal energy also changes because
of the thermalization of gravitational energy. Following S83,
we will assume that a constant fraction $\Phi$ of the gravitational
energy in bulk motion is thermalized, so that
\begin{equation}
{\mathcal{L}} \simeq A_{\rm CIE}\rho T^\alpha X f_{\rm H_2} - A_\Phi
\rho^{1/2} T
\label{cie_l_function}
\end{equation}
where
\begin{equation}
A_\Phi = {{{3k_{\rm B}}\over{2 m_H}}\over
{\sqrt{{3\pi}\over{32G}}}}\Phi \simeq
5.9\times10^4 \Phi\;{\rm erg\,cm^{1.5}\,g^{-0.5}\,K^{-1}},
\end{equation}
and we have expressed the CIE cooling rate using
eq. (\ref{approximate_cie_cool}). As we previously remarked, this last
approximation requires the gas to be predominantly molecular ($f_{\rm
H_2}\gsim0.5$); in the regions of phase space where this is not true
(that is, close to the upper temperature limit of $3000 \;{\rm K}$), the
following results are probably inaccurate.

\bfig
\psfig{figure=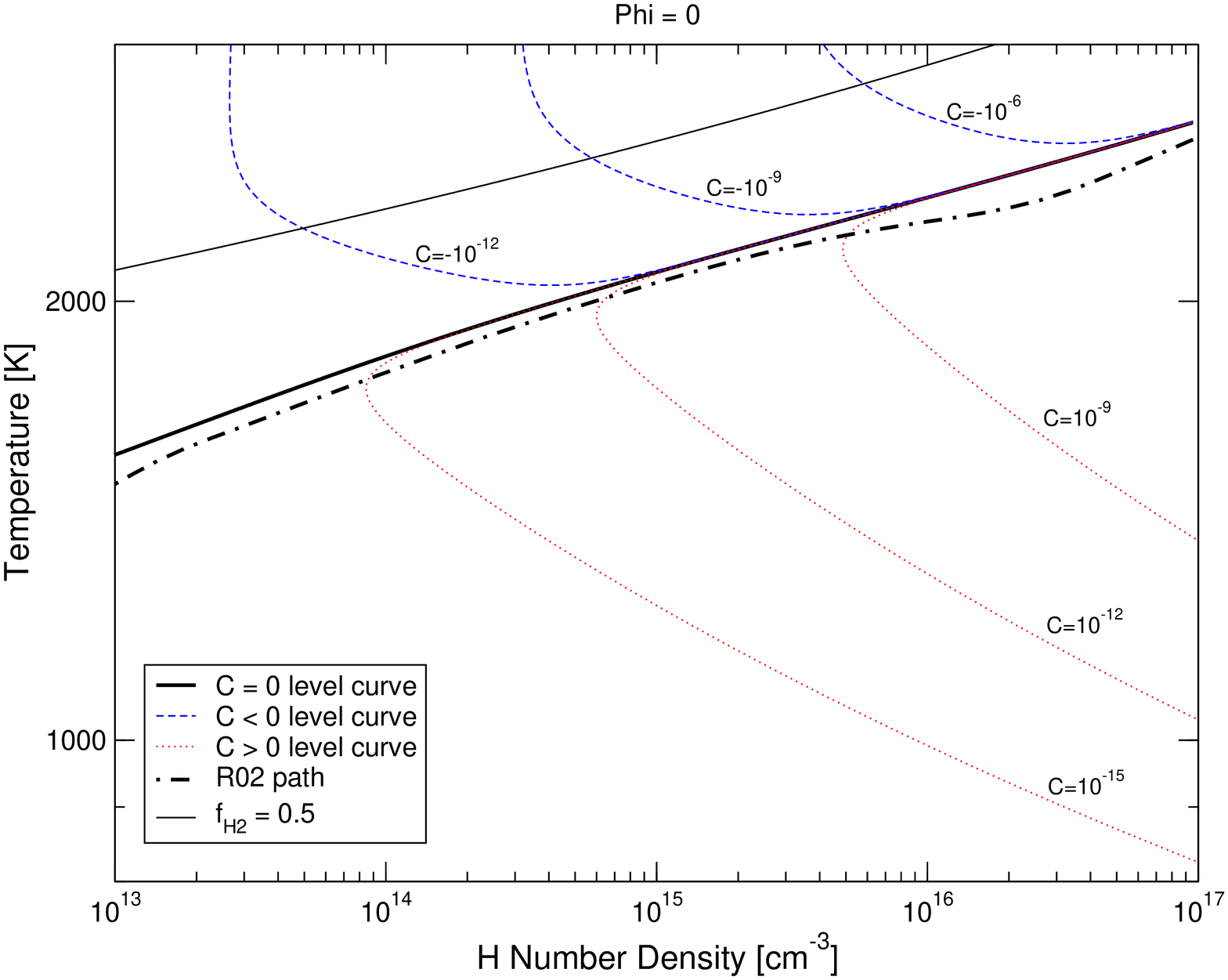,height=10truecm,angle=270}
\caption{Value of $C$ assuming no conversion of gravitational into
thermal energy ($\Phi=0$). We show the $n-T$ plane with the dashed thin
contours in the unstable region corresponding to $C=-10^{-12}, -10^{-9}$
and $-10^{-6}\;{\rm s^{-2}}$, the solid thick line for $C=0$, and the
thin dotted contours with $C=10^{-15}, 10^{-12}$ and $10^{-9}\;{\rm
s^{-2}}$ in the stable part of the phase diagram. The thick dot-dashed
line marks the ``R02 path'' (the path followed by the central regions of
the primordial proto--stellar cloud simulated in R02). The thin solid
line, corresponding to $f_{\rm H_2}(\rho,T)=f_0(\rho,T)=0.5$, separates
the low temperature region where H is mostly molecular (and
eq. \ref{approximate_cie_cool} can be used) from the high temperature
region where H is mostly atomic.}
\label{c_plot0}
\efig

\subsubsection{The instability region}
We examine the sign of $C$ across the $n-T$ phase space by assuming that
at each point we have an equilibrium \HH abundance (that is, $f_{\rm
H_2}(\rho,T)=f_0(\rho,T)$) and that the values of ${\mathcal F}$ and
${\mathcal L}$ (and their derivatives) can be found through
eqs. (\ref{molecular_formation_rate}) and (\ref{cie_l_function}),
respectively (see the appendix for the explicit expressions of the
derivatives). We still need to assume a value for the $\Phi$ parameter
in eq. (\ref{cie_l_function}),
and we choose to investigate two of the most relevant values: $\Phi=0$
and $\Phi=0.5$. The former is consistent with the assumed spherical
collapse scenario, while the latter is the value that can be expected if
a disk in keplerian rotation forms, although such scenario is not
consistent with the rest of our assumptions (for example, in a disk
geometry the cooling rate is likely to be significantly higher).

In Figures \ref{c_plot0} and \ref{c_plot1}, we show contour plots of the
values of $C$ in the $n-T$ plane in the two cases.

In both plots it can be seen that at high temperatures
($T\gsim1500-2500\;{\rm K}$, depending on $n$ and $\Phi$) there is a
region where $C<0$ and the chemo-thermal instability can occur.  Since
the conversion of gravitational energy into thermal energy has a
stabilizing effect, the size of the unstable region keeps reducing with
increasing values of $\Phi$, but it never disappears, even when $\Phi=1$.


 
\bfig
\psfig{figure=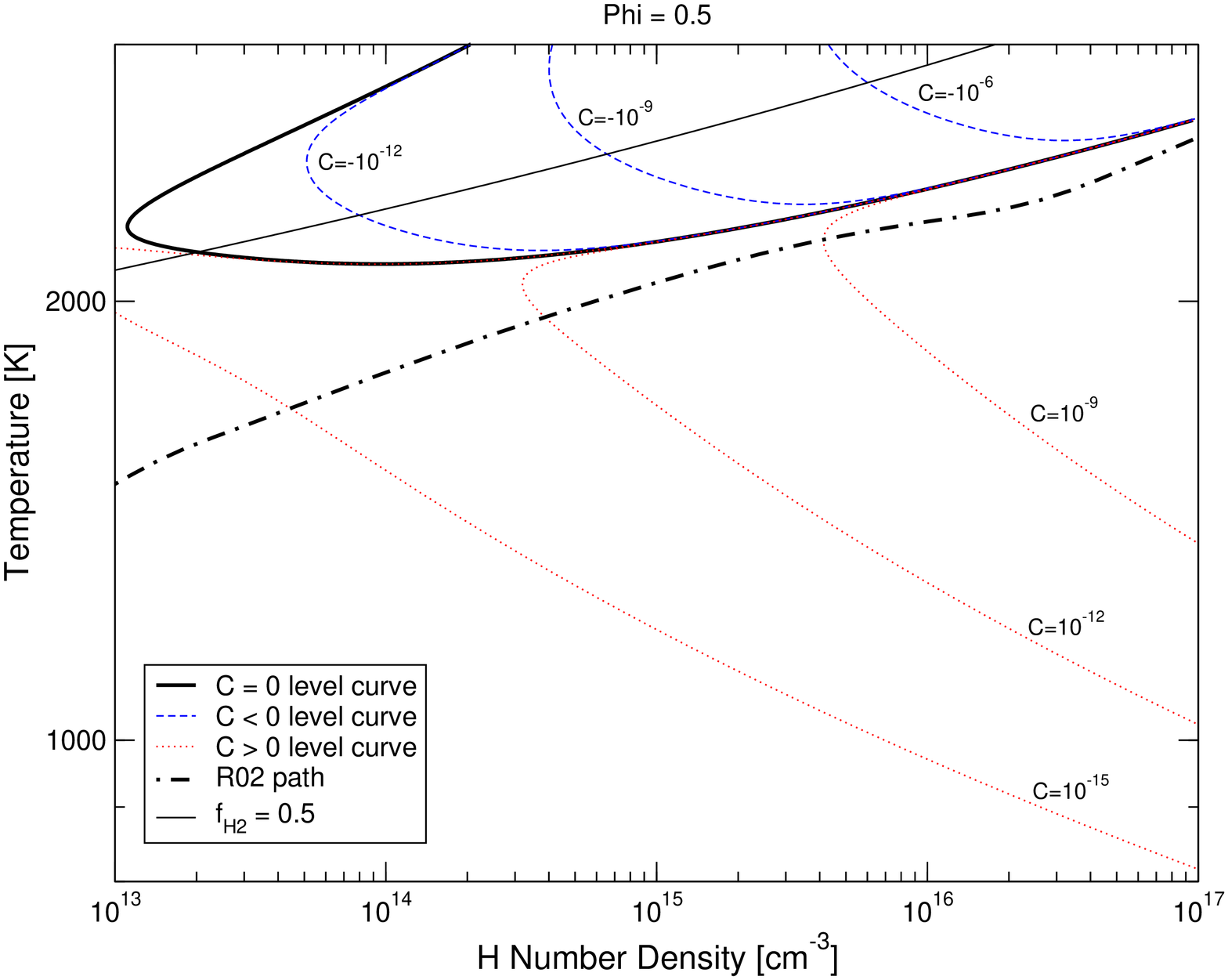,height=10truecm,angle=270}
\caption{Same as Fig.~\ref{c_plot0}, assuming a 50\%
conversion of gravitational into thermal energy ($\Phi=0.5$).}
\label{c_plot1}
\efig

\subsection{Instability on the R02 path}

In figures \ref{c_plot0} and \ref{c_plot1} it is quite clear that the
R02 path always remains outside the unstable region ($\Phi=0.5$), or
barely grazes it ($\Phi=0$).

However, this results suffers from some important uncertainties: first
of all, we already mentioned that the third addendum in
eq. (\ref{instability_criterion}) introduces a strong dependency of
$C$ on the exact value of the \HH abundance $f_{\rm H_2}$ (and of its
time derivative ${\mathcal F}$); less importantly, it only considers
optically thin continuum cooling, but this is not completely appropriate
at the two extremes of the considered density range.

Despite these subtleties the simple fact that the numerical results of
R02 fall so close to the $C=0$ lines indicates that the presented
analysis may be also applied to the collapsing proto-stellar cloud as a
whole. Being able to describe the exact evolution and formation of the
primordial proto-star purely analytically is highly desirable and will
be the subject of a subsequent paper.

A better investigation can be accomplished by applying the
same criteria described in the previous subsections to the analysis of
the behaviour of the object simulated by R02 (and also by ON98). Since
we have access to the full evolution of such object, we can study its
properties in greater detail, and in a larger range of densities
($10^8\;{\rm cm^{-3}}\leq n_p \leq 10^{17}\;{\rm cm^{-3}}$).

First of all, we can try to fix the problems about \HH abundance. This
can be done in at least two different ways:
\begin{enumerate}
\item{take $f_{\rm H_2}$ from R02 data (rather than by assuming
$f_{\rm H_2}=f_0$) and leave everything else unchanged: in particular,
we take the values of ${\mathcal{F}}$ and of its derivatives from
eq. (\ref{molecular_formation_rate});}
\item{take $f_{\rm H_2}$ from R02 data, and also estimate values of
${\mathcal F}$ which are consistent with these data, but keep
calculating the values of the partial derivatives
${\mathcal F}_T$, ${\mathcal F}_\rho$ and ${\mathcal F}_f$ from the
analytical formulae we were using in the previous subsection.}
\end{enumerate}

Approach (i) is the simplest, and it is very useful at low density,
where the R02 abundances are very different from the equilibrium ones
because the collapsing object had not reached chemical equilibrium yet,
and where both ways of estimating ${\mathcal F}$ (from
eq. \ref{molecular_formation_rate}\ and from R02 data) lead to identical
results; on the other hand, at medium and high densities the results are
very different (despite the small difference in the $f_{\rm H_2}$
values, generally just a few percent), and it must be remarked that
there is a logical inconsistency in taking the \HH abundance and its
rate of variation from two different sources.  From this point of view,
method (ii) is clearly superior, even if not optimal: from the
purely logical point of view, we should obtain also the values of
${\mathcal F}_T$, ${\mathcal F}_\rho$ and ${\mathcal F}_f$ from the R02
data, but unfortunately there is no way in which we can do that.

Fixing the luminosity term is much simpler, since we don't have so many
uncertainties: we add a term for the \HH line luminosity into the
definition of the function ${\mathcal{L}}$, and modify the CIE cooling
rate by introducing a ``CIE optical depth'' correction factor. For the
first task, we can employ the simple equation
(\ref{powerlaw_thick_cool}) combined with the Hollenbach \& McKee (1979)
cooling rate, while for the second we have empirically found that a
correction term of the form ${{1-e^{-\tau_c}}\over{\tau_c}}$ (with
$\tau_c=(\rho/\rho_{0,c})^{2.8}$ and $\rho_{0,c}=3.3\times10^{-8}\; {\rm
g\,cm^{-3}}$), in conjunction with a slight reduction of the
normalization ($A_{\rm CIE,R02}=0.054\;{\rm
erg\,cm^3\,g^{-2}\,s^{-1}\,K^{-4}}$, the difference is likely due to the
different data set used by R02 for estimating CIE cooling) provides an
excellent fit to the R02 CIE data up to the highest densities we are
considering: the agreement with R02 data is actually good at least up
to $n\gsim 10^{19}\; {\rm cm^{-3}}$ (figures \ref{evolution_comparison}
and \ref{structure_comparison} were obtained with this approximation).
Finally, we will only consider the $\Phi=0$ case, so in the following we
assume that

\begin{equation}
{\mathcal{L}} =
{{Xf_{\rm H_2}}\over{m_{\rm H}}}{\mathcal{H}}
\max{[1,(\rho/\rho_{0,l})^{-\beta_l}]} +
A_{\rm CIE,R02} T^\alpha X \rho f_{\rm H_2}
{{1-e^{-\tau_c}}\over{\tau_c}}
\label{r02path_l_function}
\end{equation}

The results are shown in Fig. \ref{r02_plot}, where we show the
evolution of the value of the $C$ parameter and of the corresponding
instability growth timescale along the R02 path, as a function of
density. The two sets of panels show the results obtained with each of
the above assumptions about the values of $f_{\rm H_2}$ and
${\mathcal{F}}$.

\bfig
\psfig{figure=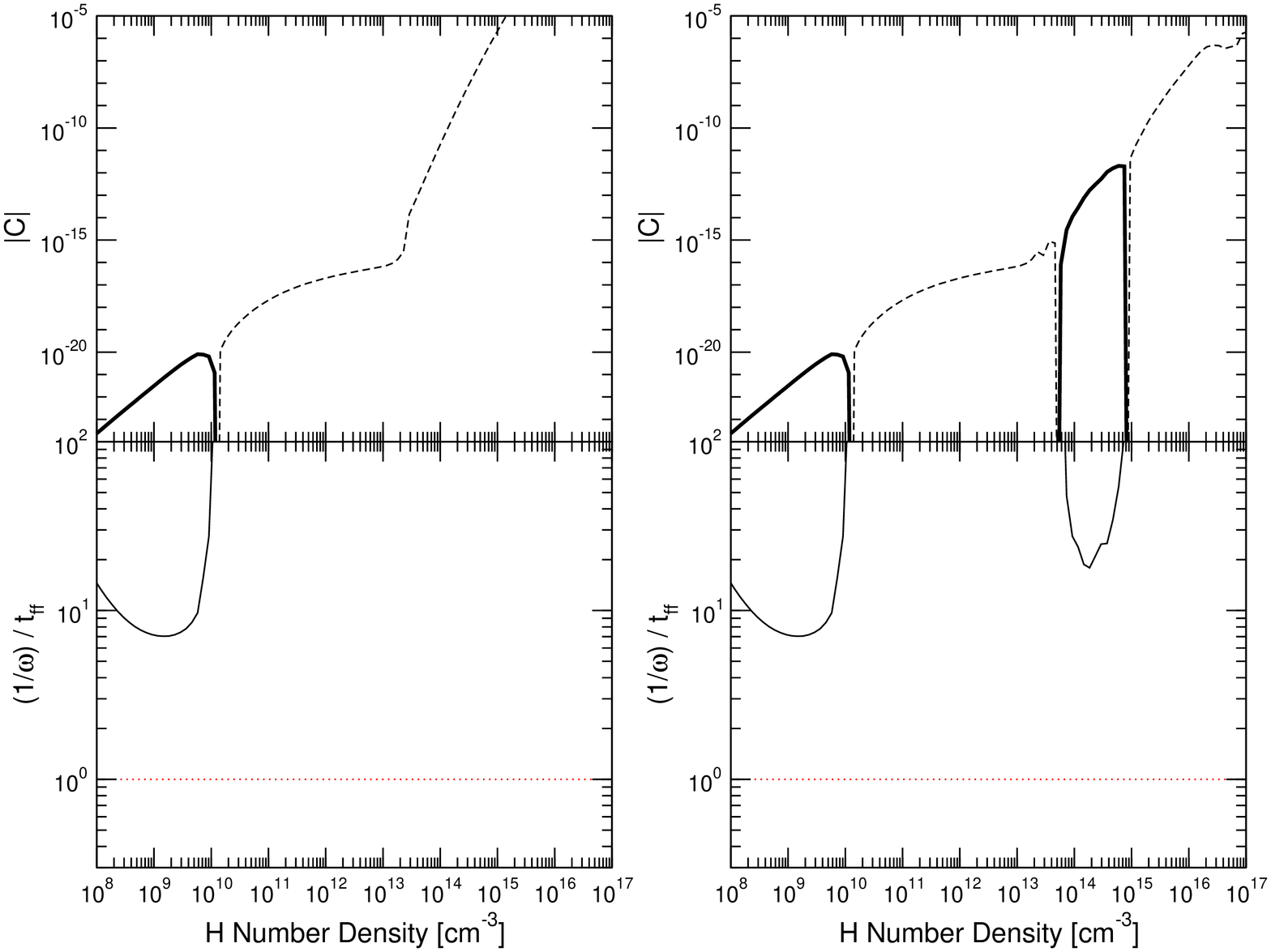,width=16.5truecm,angle=270}
\caption{Value of the parameter $C$ (top panels) and of the ratio of the
fluctuation growth timescale $1/\omega$ to the free-fall timescale
$t_{\rm ff}$ (bottom panels) along the R02 path, with two different
assumptions about the \HH fraction variation rate ${\mathcal{F}}$. In
the left panels we take the \HH fraction $f_{\rm H_2}$ from the R02
results, but use the analytical values of ${\mathcal{F}}$ (case i). In
the right panels we use the R02 data to get both $f_{\rm H_2}$ and
${\mathcal{F}}$, but we continue to estimate the derivatives of
${\mathcal{F}}$ from analytical formulae (case ii). In the upper panels,
the ranges where $C<0$ (which implies instability) are denoted through a
thick line, while thin dashed lines denote the ranges where $C>0$ (and
the object is chemo-thermally stable).  In the bottom panels we do not
show this ratio when $C>0$ since that implies $\omega<0$.}
\label{r02_plot}
\efig

First of all, it is apparent that the R02 object is unstable at low
densities ($n\lsim2\times10^{10}\;{\rm cm^{-3}}$), and this is
independent from the assumptions about ${\mathcal{F}}$. This regime,
where the instability is due to the fast \HH formation at the onset of
3-body reactions, corresponds to the one found by PSS83 and S83, and
recently discussed by ABN02 and OY03. However, the bottom panels clearly
show that the instability growth timescale is always significantly
longer than the free-fall timescale, so that this kind of instability
cannot actually lead to fragmentation. This result is slightly different
from the one recently obtained by OY03, who find that there exists a
density range about $n\simeq5\times10^9\;{\rm cm^{-3}}$ where the growth
timescale is slightly shorter than the free fall one, but their
conclusion (the instability is present but does not lead to
fragmentation) is the same as ours\footnote{This difference could
partially arise from the slightly different definition of dynamical
timescale used by OY03; see their eqn. A16.}. A different
interpretation for the lack of fragmentation is the one pointed out by
ABN02, that is, mixing due to turbulent motions: if these turbulent
motions move at about the speed of sound $c_s$, they should be able to
erase any fluctuation of size $\lsim c_s/\omega$, and we find that the
``core'' where fluctuations could develop is always several times
smaller than that.

The second result is that CIE cooling can lead to an instability, at
least if we estimate ${\mathcal{F}}$ from the R02 data (second
approach): the instability occurs over about one order of magnitude in
density, when continuum optically thin cooling is dominant. However,
the bottom panels show that a situation similar to the one described
at low density is very likely to occur: the instability growth
timescale is always much longer than the free fall timescale, and
fluctuations cannot be large enough to survive turbulent mixing. We
also conducted some experiments taking reasonable guesses at the
values of ${\mathcal{F}}_T$, ${\mathcal{F}}_\rho$ and
${\mathcal{F}}_f$, with the result that the unstable range could
extend over the whole phase when the cooling is due to optically thin
continuum emission, but the fluctuation growth is always too slow to
produce fragmentation.

\section{Discussion and Conclusions}
We have investigated the chemical and thermal instabilities of
primordial metal-free gas during its collapse towards the formation of a
protostar. We have focused on the Collision Induced Emission (CIE) and
on its effects on fragmentation. We point out that even when \HH line
cooling becomes optically thick at densities $n\gsim10^{10}\;{\rm
cm^{-3}}$ this does not necessarily set a minimum mass scale for
fragmentation due to an opacity limit. At higher densities
($n\gsim10^{14}\;{\rm cm^{-3}}$) the optically thin continuum (CIE)
cooling dominates, and the collapsing protostellar cloud can once again
fulfill the necessary (but not sufficient) fragmentation criterion
$t_{\rm cool}<t_{\rm dyn}$.  We present accurate approximations to
cooling and chemical rates that are particularily useful for analytical
studies. A detailed analysis of the chemo--thermal stability of
collapsing primordial gas leads us to conclude that the thermal
instability arising from optically thin continuum cooling is of similar
strength as the three body H$_2$ formation instability discussed
previously (PSS83; S83; OY03). However, our analytical analysis of the
low density instability confirms the full 3D simulations results of Abel
et al.~(2002) in showing that formally the instability is very strong
yet it cannot grow sufficiently fast to lead to independently collapsing
fragments.  In hindsight, it may not be too surprising that this
instability does not lead to fragmentation. It fulfills the necessary
(but not sufficient) condition of leading to a cooling time shorter than
the dynamical time. An unstable patch of gas will, however, grow
initially only iso--barically as it is compressed from the surrounding
higher pressure regions. So although the cooling rate can increase
dramatically as one increases the H$_2$ fraction by a factor one
thousand the minimum temperature the gas cools to cannot.  Consequently,
correspondingly small density fluctuations are formed which do not
become gravitationally unstable and become mixed back in into the
slightly warmer surrounding regions.

Given the efficiency of collision induced emission cooling
(Figure~\ref{cooling_comparison}) and its relatively small effect on the
temperature in the one dimensional results one is tempted to conclude
that it will also not lead to fragmentation. This is most likely a fair
assessment since again the growth times are long and the initial
iso--baric growth leads to only small overdensities. However, in a fully
three--dimensional calculation the CIE cooling might be able to cool a
disk forming around the primordial proto--star sufficiently far in order
to become gravitationally unstable and form stellar, or even planetary
size companions (see e.g. Boss 1993 and Boss 2002). Given the dramatic
accretion rates of these proto--stars, however, it is not clear whether
a sufficiently massive disk may form in the first place preventing a
conclusive answer.

Despite the simplicity of our derivation the analytical rates for the
\HH line cooling rate in the optically thick regime and optically
thick continuum cooling they are in remarkably good agreement with the
radiative transfer calculations of R02 and ON98 that followed the
detailed transfer of hundreds of molecular hydrogen rotational and
vibrational lines.  We have shown that implementing these simple
approximations into the one dimensional Lagrangian hydrodynamics code
developed by R02 leads to virtually
indistinguishable results over all relevant density and temperature
ranges. This clearly demonstrates that at least in the initial phases
of primordial proto--stellar formation the effects of radiative
transfer are in essence a local correction to cooling rather than a
transport of energy to distant regions of the hydrodynamic flow. This
is an important result which, given the analytical and purely local
cooling correction factors derived here should one allow to follow the
proto--stellar collapse in three dimensions all the way to stellar
densities employing the recently developed versions of 128bit adaptive
mesh refinement techniques of Bryan, Abel \& Norman (2001) or similar
high dynamic range hydrodynamic codes.

\section*{Aknowledgements}
We thank K. Omukai for useful discussions and for providing us
preliminary version of his OY03 paper. This work has been supported
in part by the NSF CAREER award-\#AST-0239709.

\appendix
\section{Summary of Formulae and Derivatives}

\subsection{Luminosity}

The luminosity per unit mass ${\mathcal{L}}$ can be written as (see
eq. \ref{r02path_l_function})
\begin{equation}
{\mathcal{L}} \simeq
{{Xf_{\rm H_2}}\over{m_{\rm H}}}{\mathcal{H}}
\max{[1,(\rho/\rho_{0,l})^{-\beta_l}]} +
A_{\rm CIE,R02} T^\alpha X \rho f_{\rm H_2}
{{1-e^{-\tau_c}}\over{\tau_c}}
- A_\Phi T \rho^{1/2},
\end{equation}
where we have included the thermalization of gravitational energy term.
The constants values are $\rho_{0,l}=1.34\times10^{-14}\;{\rm g\,cm^{-3}}$,
$\beta_l=0.45$, $A_{\rm CIE,R02}=0.054\;{\rm erg\,cm^3\,g^{-2}\,K^{-4}}$,
$\alpha=4$, $\tau_c=(\rho/\rho_{0,c})^{2.8}$
($\rho_{0,c}=3.3\times10^{-8}\;{\rm g\,cm^{-3}}$) and
$A_\Phi=5.9\times10^4\;{\rm erg\,cm^{1.5}\,g^{-0.5}\,K^{-1}}$;
${\mathcal{H}}(T_3)$ is given by eq. (\ref{lte_thin_lines_cooling_ht}).

If the collapsing object remains approximately spherical, this formula
is reasonably accurate up to densities $\rho\simeq2\times10^{-5}\;{\rm g\,
cm^{-3}}$.

Because of the maximum in the \HH lines term, it is useful to
distinguish two cases.

if $\rho\leq\rho_{0,l}$, the partial derivatives of ${\mathcal{L}}$ are
\begin{eqnarray}
{\mathcal{L}}_T = &
{{\partial{\mathcal{L}}}\over{\partial T}} = &
{{Xf_{\rm H_2}}\over{m_{\rm H}}}{{\partial{\mathcal{H}}}\over{\partial T}} +
\alpha A_{\rm CIE} T^{\alpha-1} X \rho f_{\rm H_2}
{{1-e^{-\tau_c}}\over{\tau_c}} - A_\Phi \rho^{1/2}\\
{\mathcal{L}}_\rho = &
{{\partial{\mathcal{L}}}\over{\partial \rho}} = &
A_{\rm CIE} T^\alpha X f_{\rm H_2}
\left[{\beta_c e^{-\tau_c}+(1-\beta_c){{1-e^{-\tau_c}}\over{\tau_c}}}\right]-
{1\over2} A_\Phi T \rho^{-1/2}\\
{\mathcal{L}}_f = &
{{\partial{\mathcal{L}}}\over{\partial f_{\rm H_2}}} = &
{{X}\over{m_{\rm H}}}{\mathcal{H}} + A_{\rm CIE} T^\alpha X \rho
{{1-e^{-\tau_c}}\over{\tau_c}}
\end{eqnarray}

Instead, if $\rho_{0,l}\leq\rho$ we have:
\begin{eqnarray}
{\mathcal{L}}_T = &
{{Xf_{\rm H_2}}\over{m_{\rm H}}}{{\partial{\mathcal{H}}}\over{\partial T}} (\rho/\rho_{0,l})^{-\beta_l} +
\alpha A_{\rm CIE} T^{\alpha-1} X \rho f_{\rm H_2}
{{1-e^{-\tau_c}}\over{\tau_c}} -
A_\Phi \rho^{1/2}\\
{\mathcal{L}}_\rho = &
-{{\beta_l}\over{\rho_{0,l}}} {{Xf_{\rm H_2}}\over{m_{\rm H}}}{\mathcal{H}}
(\rho/\rho_{0,l})^{-(\beta_l+1)} +
A_{\rm CIE} T^\alpha X f_{\rm H_2}
\left[{\beta_c e^{-\tau_c}+(1-\beta_c){{1-e^{-\tau_c}}\over{\tau_c}}}\right]-
{1\over2} A_\Phi T \rho^{-1/2}\\
{\mathcal{L}}_f = &
{{X}\over{m_{\rm H}}}{\mathcal{H}} (\rho/\rho_{0,l})^{-\beta_l} +
A_{\rm CIE} T^\alpha X \rho {{1-e^{-\tau_c}}\over{\tau_c}}
\end{eqnarray}

where, in both cases, 

\begin{eqnarray}
{{\partial{\mathcal{H}}}\over{\partial T}} =
{{\partial T_3}\over{\partial T}}
{{\partial{\mathcal{H}}}\over{\partial T_3}} =
{{10^{-3}}\over{T_3^2}} \left[{
\left({{9.5\times10^{-22}T_3^{3.76}}\over{1+0.12T_3^{2.1}}}\right)
e^{-\left({{0.13}\over{T_3}}\right)^3}
\left({{{3.76T_3+0.2T_3^{3.1}}\over{1+0.12T_3^{2.1}}}+
{{0.0066}\over{T_3^2}}}\right)}\right] + \nonumber\\
+ {{10^{-3}}\over{T_3^2}} \left[{
1.5\times10^{-24}e^{-{{0.51}\over{T_3}}} +
3.9\times10^{-18}e^{-{{5.86}\over{T_3}}} +
1.9\times10^{-17}e^{-{{11.7}\over{T_3}}}}\right]
\quad {\rm erg\,s^{-1}\,K^{-1}}
\end{eqnarray}

\subsection{\HH Formation}
The \HH formation function is given by

\begin{equation}
{\mathcal{F}} = {{\partial{f_{\rm H_2}}}\over{\partial t}} =
n_f [2n_{\rm p}k_4 (1-f_{\rm H_2})^2 - k_5 f_{\rm H_2}]
\end{equation}
where
\begin{equation}
n_{\rm p} \equiv \rho X /m_{\rm H}, \qquad
n_f = n_{\rm p}(1-15f_{\rm H_2}/16)
\end{equation}
and (as explained in section 3.2.2)
\begin{equation}
k_4 = A_4 T^{-1}, \qquad k_5 = A_5 T^{0.2} e^{-B_5/T} (1-e^{-C_5/T}),
\end{equation}
with $A_4=5.5\times10^{-29}\;{\rm cm^6\,s^{-1}}$,
$A_5=2.2\times10^{-9}\;{\rm cm^3\,s^{-1}}$, $B_5=51800\;{\rm K}$,
$C_5=6000\;{\rm K}$.

The partial derivatives are
\begin{eqnarray}
{\mathcal{F}}_T \equiv & 
{{\partial{\mathcal{F}}}\over{\partial T}} = &
-{{n_f}\over T} \left[{2n_{\rm p}k_4(1-f_{\rm H_2})^2 +
k_5f_{\rm H_2} \left({0.2 + {{B_5}\over T} +
{{C_5}\over T}{1\over{e^{C_5/T}-1}}}\right)}\right]\\
{\mathcal{F}}_\rho \equiv &
{{\partial{\mathcal{F}}}\over{\partial \rho}} = &
{{n_f}\over\rho}[4n_{\rm p}k_4(1-f_{\rm H_2})^2 - k_5f_{\rm H_2}]\\
{\mathcal{F}}_f \equiv &
{{\partial{\mathcal{F}}}\over{\partial f_{\rm H_2}}} = &
-n_f \left\{{{{15}\over{16-15f}}
[2n_{\rm p}k_4(1-f_{\rm H_2})^2-k_5f_{\rm H_2}] +
4n_{\rm p}k_4(1-f_{\rm H_2})+k_5}\right\}
\end{eqnarray}

\label{lastpage}


\begin{thebibliography}{99}
\bibitem{}Abel T., Anninos P., Norman M.L., Zhang Y., 1998, ApJ, 508, 518
\bibitem{}Abel T., Bryan G.L., Norman M.L., 2000, ApJ 540, 39
\bibitem{}Abel T., Bryan G.L., Norman M.L., 2002, Science, 295, 93 [ABN02]
\bibitem{}Bonnor W.B., 1956, MNRAS, 116, 351
\bibitem{}Borysow A., Jorgensen U.G., Fu Y., 2001, JQSRT, 68, 235
\bibitem{}Borysow A., 2002, A\&A, 390, 779
\bibitem{}Boss A.P., 1993, ApJ, 410, 157
\bibitem{}Boss A.P., 2002, ApJ, 567, L149
\bibitem{}Bromm V., Coppi P.S., Larson R.B., 2002, ApJ, 564, 23
\bibitem{}Bryan G.L., Abel T., Norman M.L., 2001, in Proc. ACM/IEEE
	Conf. on Supercomputing (CD-ROM; New York: ACM Press), 13
\bibitem{}Couchman, H.~M.~P.~\& Rees, M.~J.\ 1986, MNRAS, 221, 53
\bibitem{}Ebert R., 1955, Zs. Ap., 217
\bibitem{}Frommhold L., 1993, Collision-induced absorption in gases, Cambridge University Press, Cambridge (UK)
\bibitem{}Fuller T.M., Couchman H.M.P., 2000, ApJ, 544, 6
\bibitem{}Galli D., Palla F., 1998, A\&A, 335, 403
\bibitem{}Gustafsson M., Frommhold L., 2001, ApJ 546, 1168\
\bibitem{}Gustafsson M., Frommhold L., Meyer W., 2003, Journal of
Chemical Physics, 118, 1667
\bibitem{}Jorgensen U.G., Hammer D., Borysow A., Falkesgaard J., 2000, A\&A, 361, 283 
\bibitem{}Larson R.B., 1969, MNRAS, 145, 271
\bibitem{}Lenzuni P., Chernoff D.F., Salpeter E.E., 1991, ApJS, 76, 759
\bibitem{}Lepp S., Shull J.M., 1983, ApJ, 270, 578
\bibitem{}Machacek M.E., Bryan G.L., Abel T., 2001, ApJ, 548, 509
\bibitem{}Martin P.G., Schwarz D.H., Mandy M.E., 1996, ApJ, 461, 265
\bibitem{}Nishi R., Susa H., 1999, ApJ, 523, 103
\bibitem{}Omukai K., Nishi R., 1998, ApJ, 508, 141 [ON98]
\bibitem{}Omukai K., Yoshii Y., 2003, ApJ, accepted (astro-ph/0308514) [OY03]
\bibitem{}Palla F., Salpeter E.E., Stahler S.W., 1983, ApJ, 271, 632
\bibitem{}Penston M.V., 1969, MNRAS, 144, 425
\bibitem{}Rees M.J., 1976, MNRAS, 176, 483
\bibitem{}Ripamonti E., Haardt F., Ferrara A., Colpi M., 2002, MNRAS 334, 401 [R02]
\bibitem{}Sabano Y., Yoshii Y., 1977, PASJ, 29, 207
\bibitem{}Saio H., Yoshii Y., 1986, ApJ, 301, 587
\bibitem{}Silk J., 1983, MNRAS, 205, 705 [S83]
\bibitem{}Tegmark M., Silk J., Rees M., Blanchard A., Abel T., Palla F., 1997, ApJ, 474, 1
\end{thebibliography}
\end{document}